\newif\ifanonymized
\anonymizedfalse    

\newif\ifsubmit
\submitfalse	

\documentclass[acmlarge]{acmart}
\usepackage{tabularx}
\usepackage{placeins}
\usepackage{tcolorbox}
\usepackage{xcolor}
\definecolor{summary}{HTML}{529077}
\definecolor{message}{HTML}{F09654}
\definecolor{researcher}{HTML}{838383}
\definecolor{error}{HTML}{C9574B}
\usepackage{listings}
\usepackage{fancyhdr}

\usepackage{url}

\usepackage{color}

\usepackage[mmddyyyy,hhmmss]{datetime}

\usepackage{xspace}

\usepackage{graphicx}
\DeclareGraphicsExtensions{.pdf,.png,.eps,.ps,.jpg}

\ifsubmit
  \newcommand{\hey}[1]{\relax}
  \newcommand{\heyvarun}[1]{\relax}
  \newcommand{\bibtex}[1]{\relax}
  \newcommand{\note}[1]{\relax}
\else
  \newcommand{\hey}[1]{\textcolor{magenta}{[{#1}]}}
  \newcommand{\heyvarun}[1]{\textcolor{blue}{[Varun: {#1}]}}
  \newcommand{\note}[1]{\par\textcolor{magenta}{Note: {#1}}\par}
  \newcommand{\bibtex}[1]{\textcolor{red}{@bibtex}\{#1\}} 
\fi

\newcommand{\hide}[1]{\relax}

\newcommand{\seclabel}[1]{\label{sec:#1}}

\begin{document}

\title{An Open Book, On Whose Terms? Power Dynamics in Sensor-Derived, LLM-Generated Summaries Sharing Between Older Adults and Remote Families}
\date{}

\ifanonymized
   \author{Anonymized for blind submission}
\else
   \author{Jiachen Li}
   \affiliation{
    \institution{Northeastern University}
    \city{Boston}
    \state{MA}
    \country{USA}
    }
    
    \author{Reina Szeyi Chan}
   \affiliation{
    \institution{Northeastern University}
    \city{Boston}
    \state{MA}
    \country{USA}
    }

    \author{Shihan Fu}
   \affiliation{
    \institution{Northeastern University}
    \city{Boston}
    \state{MA}
    \country{USA}
    }

    \author{Zijian Wang}
   \affiliation{
    \institution{Northeastern University}
    \city{Boston}
    \state{MA}
    \country{USA}
    }

    \author{Akshat Choube}
   \affiliation{
    \institution{Northeastern University}
    \city{Boston}
    \state{MA}
    \country{USA}
    }

    \author{Yingtian Shi}
   \affiliation{
    \institution{Georgia Institute of Technology}
    \city{Atlanta}
    \state{GA}
    \country{USA}
    }

    \author{Rosemarie Santa Gonzalez}
   \affiliation{
    \institution{Georgia Institute of Technology}
    \city{Atlanta}
    \state{GA}
    \country{USA}
    }

    \author{Thomas Ploetz}
   \affiliation{
    \institution{Georgia Institute of Technology}
    \city{Atlanta}
    \state{GA}
    \country{USA}
    }

    \author{Agata Rozga}
   \affiliation{
    \institution{Georgia Institute of Technology}
    \city{Atlanta}
    \state{GA}
    \country{USA}
    }

    \author{Xiang Zhi Tan}
   \affiliation{
    \institution{Northeastern University}
    \city{Boston}
    \state{MA}
    \country{USA}
    }

    \author{Elizabeth Mynatt}
   \affiliation{
    \institution{Northeastern University}
    \city{Boston}
    \state{MA}
    \country{USA}
    }

    \author{Varun Mishra}
   \affiliation{
    \institution{Northeastern University}
    \city{Boston}
    \state{MA}
    \country{USA}
    }
\fi






\begin{abstract}
As older adults (OAs) increasingly favor aging in place, remote family members (RFMs) need meaningful awareness of their loved ones without intrusive monitoring. Large language models can provide rich interpretations of non-intrusive sensing data, yet recent work has largely emphasized technical accuracy over how families use or benefit from this information. We conducted formative interviews with 12 RFMs and 2 OAs, followed by a three-month deployment case study with one OA–RFM dyad. Our system provided daily AI-generated summaries of multimodal sensing data, which we iteratively refined with both parties using a research-through-design approach. Although the summaries integrated well into the dyad’s daily routine, sustaining over 90\% engagement, their lived experience revealed evolving tensions around agency, reassurance, privacy, trust, and communication, shaped by sensor limitations and existing communication misalignments. Our findings inform the design of future AI-mediated sharing systems that account for real-world sensing constraints and complex family care dynamics.

\end{abstract}
\maketitle  


\ifsubmit
    \relax
\else
\fi


\section{Introduction} 
\seclabel{introduction}




Informal caregiving for older adults (OAs) has long involved a fundamental tension: older adults want to age in place (AiP) in ways that preserve their independence and agency, while remote family members (RFMs) seek reassurance and timely awareness of their loved one's well-being from afar.
Over the past two decades, ubiquitous computing technologies have helped negotiate this tension. 
Passive sensing systems in particular, such as wearables, phones, and home sensors, have enabled RFMs to gain a window into an OA's daily life without the intrusiveness of technologies such as continuous video monitoring~\cite{alkhatib2021wants, mcneill2017functional}.
Recently, Large Language Models (LLMs) have fundamentally changed what can be done with such data. 
Rather than presenting isolated metrics such as step counts, LLMs can synthesize heterogeneous sensing streams into high-level, personalized narratives to better capture an older adult's day \cite{li2025vital,huatao2025,11206243,nepal2024contextual,stromel2024narrating,choube2025gloss}.

While most prior work using LLMs to understand passive sensing data has focused on benchmarking and improving the accuracy of the outputs~\cite{choube2025gloss,yang2024drhouse,ji2024hargpt,kim2024health,cosentino2024towards,ouyang2024llmsense,fang2024physiollm,shi2026trace}, more accurate and richer interpretation of the data does not necessarily translate into a better experience for RFMs.
Such interpretations may provide reassurance, but they may also reshape expectations, communication patterns, and perceptions of privacy between older adults and their family members. 
While existing work has largely focused on improving LLMs' ability to interpret multi-modal sensing data, far less attention has been paid to how these interpretations should be designed, negotiated, and shared within real family relationships.
A familiar tension between OAs and RFMs resurfaces in a new form: \textbf{how can these emerging AI capabilities support the sharing of OAs' daily lives with RFMs in ways that benefit both and support AiP?}


To explore this question, we followed a research-through-design approach~\cite{zimmerman2007research}, first conducting a literature review and 14 preliminary interviews with OAs and RFMs to draw high-level insights that informed the design of the summary used in the case study.
From the interviews, we found that RFMs' needs were highly personalized across different living situations, and that OAs' sense of privacy violation was difficult to capture without observing how they responded when summaries were actually sent to a loved one. 
As a result, rather than conducting more interviews, we moved forward with a longitudinal case study involving one dyad, to examine their lived experiences and evolving communication that emerged around AI-mediated sharing, aiming to surface trade-offs and design tensions that are difficult to capture through interviews.
We customized the summaries according to the dyad's needs in pre-deployment interviews, and began delivering daily summaries generated from the OA's (O1's) data to her son (F1) for three months. 
During the deployment, we iteratively refined them through weekly check-ins with both parties.

The dyad successfully integrated the summaries into their daily routine, sustaining an engagement rate of 95.6\% for OA and 90.0\% for RFM across three months, even after check-ins with researchers shifted to a more casual cadence.
Although O1 expressed a willingness to share anything in the summaries with her son during the pre-deployment interview, a closer examination of their lived experience through regular check-ins revealed multiple tensions.

First, tensions arose not only from factual errors in the summaries, but also from summaries that were factually correct yet conveyed a misleading impression of the OA's day. 
For example, a summary might describe O1 as having ``remained mostly sedentary,'' consisting mostly of sentences describing indoor activities (e.g., moving around the kitchen). 
While each of these sentences was accurate according to her watch and home sensors, O1 felt the summary wrongly portrayed her as inactive and homebound; she wanted it to instead highlight her outdoor activities, such as walking her dog. 
O1 grew especially worried about sharing this kind of misleading interpretation with her son: although she knew exactly how her day had gone, she worried about how she was being perceived by him, for example, as someone who did not go out at all.

Second, these tensions were amplified by pre-existing misalignments in how the dyad communicated. 
For instance, O1 worried that the heart-rate patterns in the summary might concern F1, while F1 himself remained unconcerned, noting that occasional low heart rate readings at night were nothing to worry about and did not warrant discussion with O1. 
Because O1 and F1 maintained a low-key communication style, O1 was often unsure how F1 would react to different information in the summaries.

\medskip
\noindent
In this paper, we unpack these tensions and make three \textbf{contributions}:
\begin{itemize}
\item  We show that factual accuracy is insufficient for AI-generated summaries of personal sensing data shared with others, and demonstrate how individually accurate statements can nevertheless produce a misleading representation of an older adult’s lived experience, particularly when uneven sensing coverage causes some aspects of daily life to be disproportionately represented.
\item We characterize how AI-mediated sharing creates tensions around representational agency and introduces new expectations and obligations, where privacy concerns arise not only from what information is disclosed, but from uncertainty about how that information will be interpreted by others.
\item We derive design implications for AI systems that translate in-the-wild sensing data into interpersonal narratives, ones that account for the complex social relationships and dynamics within which such narratives are shared.
\end{itemize}
\section{Background} 
\seclabel{background}
This section reviews prior work on informal care technologies for older adults and recent advances in applying LLMs to interpret multi-modal sensing data.
\subsection{Informal Care Technologies for Older Adults Aging in Place}
According to World Health Organization (WHO), by 2030, 1 in 6 people in the world will be aged 60 years or over~\cite{who2025ageing}.
Rather than relocating to institutional care, 90\% of older adults report a strong preference to remain in their own homes and familiar communities as they grow older~\cite{farber2011aging, institute2010}. This preference is widely described as ``Aging in Place (AiP)''~\cite{centers2013public}, emphasizing autonomy, sense of security and identity, and community connection in later life~\cite{emlet2012importance,wiles2012meaning,vasunilashorn2012aging}.
Despite its appeal, AiP complicates care coordination, as RFMs often become informal primary caregivers long before professional care or long-term care transitions occur \cite{bom2019impact,diniz2018comparative,schulz2020family,ausubel2020}.
During this time, RFMs bear significant responsibility for caring for and ensuring the well-being of their loved ones.
With limited visibility into older adults’ daily lives, however, they often struggle to monitor their well-being, respond promptly to potential accidents, and maintain connections~\cite{verbakel2018caregiving,marinho2022burden}.

To address this challenge, there is growing interest in aged care monitoring technologies that are able to provide awareness of older adults’ wellbeing to assist the care coordination process with RFMs~\cite{mynatt2001digital, rowan2005digital,farber2011aging, consolvo2004carenet, caldeira2017senior, vines2013making}. 
Visual-information-based monitoring systems like indoor cameras have been widely adopted to help adult children learn about their parents’ activities of daily living at home~\cite{wagner2012review, felzenszwalb2008discriminatively, nait2004activity,wang2024ai}. These methods, however, suffer from privacy and security issues, thus making the adoption of those systems challenging~\cite{alkhatib2021wants, mcneill2017functional}.
With the growing popularity of commercial devices such as Fitbit and Apple Watch, more families are adopting sensing technologies to support aging in place~\cite{nurain2023left,schlomann2017case,vargemidis2020wearable,mcmahon2016older,short2025tracking,steinert2018activity}, with the most common devices including phone~\cite{joe2013older,seifert2017use}, wearables (e.g. smartwatch, ring)~\cite{vargemidis2020wearable,kononova2019use,alharbi2019data,peng2021habit,Bilius_Vatavu_2023}, smart home sensors~\cite{caldeira2023compare,mynatt2001digital,mynatt2004aware,lee2015sensor,fritz2025detecting}, and conversational agent (e.g., Alexa)~\cite{even2022benefits,wargnier2015towards,zubatiy2021empowering,wolfe2025caregiving,yang2024talk2care}.
While these systems provide valuable support for monitoring older adults’ well-being, understanding how to interpret multi-modal data streams and how to present such information to key informal caregivers like RFMs remain underexplored. 
For common commercial devices and applications, data representations still largely focus on simple aggregated statistics (e.g., step count visualizations) and self-reflection~\cite{8268154}. 
Data-sharing functionalities are typically limited to basic data access, rather than offering customized views that account for different caregiving roles or support care coordination and communication.

\subsection{LLM Interpretation for Multi-modal Sensing Data}


Prior work using LLM to interpret sensing data has largely focused on tasks with well-defined evaluation metrics, such as activity recognition, stress estimation, or clinical diagnosis~\cite{yang2024drhouse,ji2024hargpt,kim2024health,cosentino2024towards,ouyang2024llmsense,fang2024physiollm,shi2026trace}. Much of this literature emphasizes benchmarking model performance, such as accuracy, efficiency, or semantic fidelity, rather than supporting real-world sensemaking for end users~\cite{choube2025gloss}. 

Recent work has further explored grounding sensor data in natural language. Sensor-language models such as LLaSA~\cite{imran2024llasa}, SensorLLM~\cite{li2025sensorllm}, and SensorLM~\cite{zhang2026sensorlm} seek to align wearable-sensor time-series data with natural-language representations, while systems such as SensorChat~\cite{yu2025sensorchat} support long-term question answering over multimodal sensor data through explicit data-querying processes.
In parallel, researchers have started to explore the potential of LLMs to generate higher-level interpretations and summaries from multi-modal sensing data, shifting attention from raw prediction to narrative understanding~\cite{li2025vital,nepal2024contextual,nepal2024mindscape}.
A prominent line of work in this space focuses on self-reflection, particularly through LLM-assisted journaling that integrates passive sensing data. 
For example, MindScape integrates behavioral signals such as conversational engagement, sleep, and location with LLMs to generate personalized journaling prompts that encourage emotional reflection and well-being~\cite{nepal2024contextual,nepal2024mindscape}. 
Similarly, systems such as DailyLLM focus on generating rich, context-aware activity logs by integrating multi-modal sensing dimensions (e.g., location, motion, environment, physiology) into structured natural language representations, with an emphasis on improving semantic richness, efficiency, and deployability~\cite{11206243}. PHIA also examined LLM agents for personalized health insight generation from longitudinal wearable data through tool-augmented analysis and numerical reasoning~\cite{merrill2026transforming}.
While these efforts highlight the promise of LLMs in synthesizing complex sensing data, they primarily target self-facing use cases, where individuals have direct lived experience and ground truth about their own activities.

More recently, a few works have begun focusing on other stakeholders, e.g., Li et al. combined visual analytics with LLM-generated summaries to support expert researchers' sensemaking of sensing data \cite{li2025vital}. 
But data-sharing scenarios beyond the self add complexity: stakeholders like family members must interpret another person's life without direct visibility, while bearing potential decision-making consequences. How LLM interpretations of multi-modal sensing data should be shared with others remains largely unexplored.
\section{Method}
We applied a human-centered, iterative process throughout the study, guided by research-through-design principles. 
The following sections describe the theoretical grounding that shaped our approach, with detailed methods and findings for the literature review, formative interviews, and case study reported separately (Section 4.1/2/3, 6.1).

\subsection{Research through Design (RtD)}
Research through Design (RtD) is a methodology in which the design and iterative refinement of an artifact serves as the primary means of generating knowledge, rather than a secondary output of a hypothesis-driven study \cite{zimmerman2007research}. 
Instead of testing a fixed system against a predetermined set of metrics, RtD treats each design decision, deployment, and subsequent revision as an opportunity to surface tensions, uncover unarticulated needs, and iteratively refine both the artifact and researchers' understanding of the problem space \cite{gaver2012should}. 
We adopt this approach because the design space for sharing AI-generated summaries between older adults and RFMs is not yet well understood.
A controlled study evaluating a fixed system against predefined criteria would risk prematurely narrowing this space before the relevant design tensions were even identified. 
Instead, by deploying an evolving system with a real OA–RFM pair over an extended period, and iteratively modifying the summary, we were able to surface concrete tensions, such as the trade-off between interpretive richness and perceived intrusiveness, that would have been difficult to anticipate or evaluate through a single-shot or quantitative study design.

Following the process for RtD work~\cite{olson2014ways,zimmerman2007research}, we first select a design problem and conduct a literature review to develop an initial framing (Section 4.1), then select, iteratively evolve, and refine an idea into a completed form (Section 4.2 and Section 5), and evaluate it based on the concerns of the specific RtD practice, lab (Section 4.3) and field (Section 6). Through interviews and deployment, we reflect on the design and evaluation work and disseminate the resulting insights, repeating these steps across multiple weeks of check-ins.

\section{Initial LLM Summaries for RFMs}
In this section, we describe how we generated the initial LLM-based summaries for the subsequent interviews. 
We conducted a literature review to identify the needs of RFMs and OAs documented in prior work, and detail our modifications to the open-source Vital Insight and GLOSS systems to generate summaries informed by these needs. We then present these sample summaries to 12 RFMs and 2 OAs in interviews and gather additional feedback on their content, which we use to finalize the design of the summaries used in the later deployment study.
\subsection{Literature Review}

Past work has emphasized that retrospective summaries for family members should remain \textbf{high-level} and \textbf{qualitative}, centers on understanding ``how is my X doing?'' at a high level~\cite{mynatt2001digital,li2023privacy}. 
Summaries should also be \textbf{context-aware}, incorporating longitudinal context (e.g., health status) and temporal context (e.g., special events)~\cite{li2023privacy,wang2024redefining,sharma2023exploring,nurain2023left}. 
Given family members’ lay backgrounds and care burden, summaries must be simple and \textbf{easy to understand}; prior work has shown that graphs and visualizations may be overly complex for many users~\cite{li2023privacy}. 
When sharing the summaries with RFMs, \textbf{trust} building is crucial and should be fostered not through simple confidence scores (e.g., 70\% probability of walking) but by strategically incorporating original data sources (e.g., watch sensors indicated walking)~\cite{li2023privacy,nurain2023left,mathur2024categorizing,so2024they}. 
In terms of the content of these summaries, prior work has identified three key categories: \textbf{movement} (e.g., steps, travel, location, and motion), \textbf{crucial activities} (e.g., routine activities, low-exertion activities, physical activities and exercise, non-exercise hobbies, and social activities), and \textbf{cognitive and mental well-being} (e.g., emotion)~\cite {li2023privacy,wang2024redefining,sharma2023exploring,nurain2023left,pashmdarfard2020assessment}. 


\subsection{Initial Summary Design for the Interviews}
\begin{figure}
    \centering
    \includegraphics[width=\linewidth]{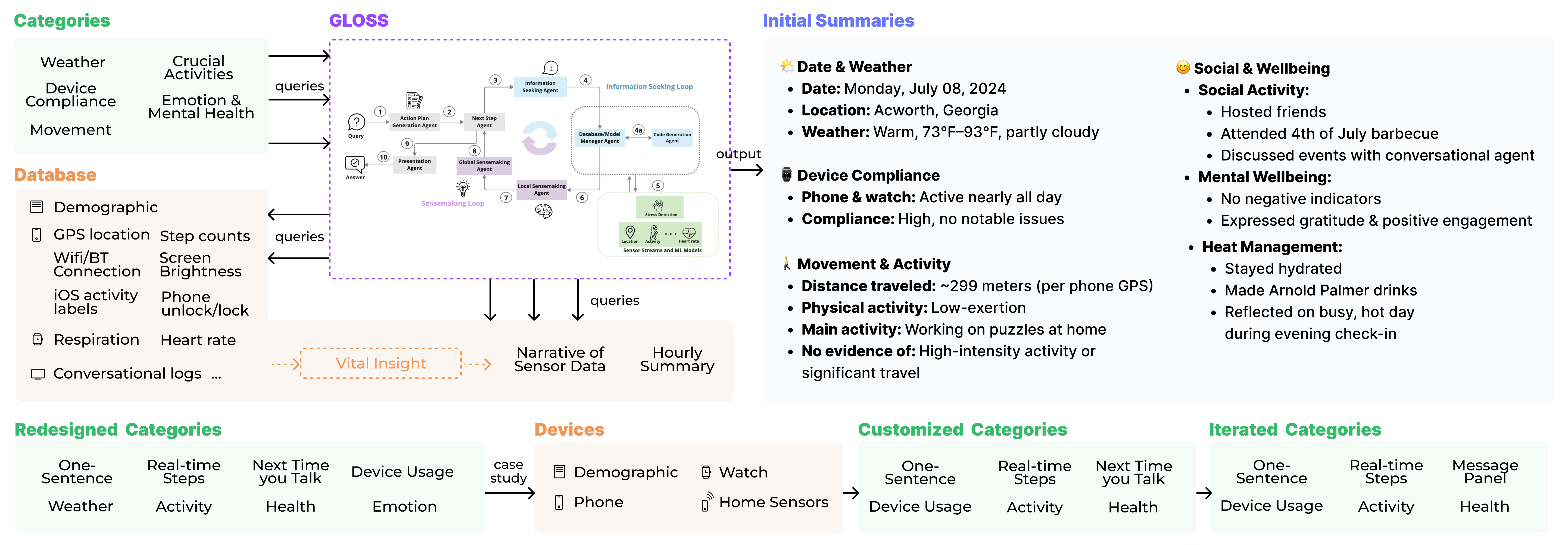}
    \caption{System architecture and sample summaries, illustrating how categories and underlying data flow through GLOSS and VI to produce the initial summaries, and how the summaries evolved from interviews, to redesign, to case-study customization, to iterative refinement during deployment. See Figures~\ref{fig:checkin1}--\ref{fig:checkin5} for summaries sent in the real-world case study.}
    \label{fig:initial}
    \Description{System architecture diagram with two rows. Top row shows the pipeline for generating initial summaries: Categories (Weather, Crucial Activities, Device Compliance, Emotion \& Mental Health, Movement) send queries to GLOSS, which sends queries to a Database. GLOSS is shown via a system diagram from prior work: a Query triggers an Action Plan Generation Agent, which passes to a Next Step Agent, which can route to an Information Seeking Loop (an Information Seeking Agent querying a Database/Model Manager Agent and a Code Generation Agent) or to a Sensemaking Loop (a Local Sensemaking Agent drawing on Sensor Streams and ML Models for stress detection, location, activity, and heart rate, feeding a Global Sensemaking Agent), before reaching a Presentation Agent that produces the Answer. The Database holds raw data (Demographic, GPS location, Wifi/BT Connection, iOS activity labels, Respiration, Step counts, Screen Brightness, Phone unlock/lock, Heart rate, Conversational logs) and Vital Insight outputs derived from that raw data (Narrative of Sensor Data, Hourly Summary). GLOSS produces a sample Initial Summary shown at right, with the following content: Date \& Weather: Date: Monday, July 08, 2024. Location: Acworth, Georgia. Weather: Warm, 73°F–93°F, partly cloudy. Device Compliance: Phone \& watch: Active nearly all day. Compliance: High, no notable issues. Movement \& Activity: Distance traveled: ~299 meters (per phone GPS). Physical activity: Low-exertion. Main activity: Working on puzzles at home. No evidence of: High-intensity activity or significant travel. Social \& Wellbeing: Social Activity — Hosted friends; Attended 4th of July barbecue; Discussed events with conversational agent. Mental Wellbeing — No negative indicators; Expressed gratitude \& positive engagement. Heat Management — Stayed hydrated; Made Arnold Palmer drinks; Reflected on busy, hot day during evening check-in. The bottom row shows how summary categories evolved across four stages: Redesigned Categories (One-Sentence, Real-time Steps, Next Time You Talk, Device Usage, each paired with Weather, Activity, Health, Emotion) branches by Devices (Demographic, Phone, Watch, Home Sensors) into Customized Categories (One-Sentence, Real-time Steps, Next Time You Talk, Device Usage paired with Activity, Health), which in turn evolves via case-study use into Iterated Categories (One-Sentence, Real-time Steps, Message Panel paired with Device Usage, Activity, Health).}
\end{figure}
To generate the sample summaries, we built on prior work on Vital Insight (VI), designed to produce expert-oriented dashboards and summaries for studies involving multi-modal personal sensing data~\cite{li2025vital}, combined with Group of LLMs for Open-ended Sensemaking (GLOSS), a query-based multi-agent system designed for open-ended sensemaking for non-expert users~\cite{choube2025gloss} with OpenAI GPT-4o (Figure \ref{fig:initial}).

For VI, we adopted its multi-level summarization pipeline, which transforms raw sensing data from different modalities into narrative sentences and then generates progressively aggregated summaries. 
For example, an LLM-generated hourly summary from 6–7AM would incorporate all modalities available during that period, including sensing data (e.g., heart rate) and contextual data (e.g., conversational agent chat logs). Each data point is first converted into a semantic representation (e.g., translating \texttt{\{wifi\_connection: 1\}} into \emph{``The phone was connected to Wi-Fi from \textit{TIME1} to \textit{TIME2}.''}), then the resulting narratives are aligned chronologically and provided to the LLM to generate an hourly summary.

Because VI was originally designed for expert audiences, we further customized the summaries for RFMs using GLOSS's multi-agent framework to tailor both content and tone. 
GLOSS supports open-ended sensemaking of passive sensing data directly from databases across multiple sensor streams, using a network of LLM agents in response to a user query. 
It requires no model training or fine-tuning, allowing it to be extended to new sensor streams with relative ease, and in evaluation substantially outperformed a standard Retrieval-Augmented Generation (RAG) baseline (87.93\% accuracy and 66.19\% consistency vs. RAG's 29.31\% accuracy and 52.85\% consistency).
Drawing on our literature review, we identified three key categories of information: movement, crucial daily activities, and cognitive and emotional wellbeing. 
We further added two categories, weather and device compliance, to surface additional context for interpreting the summary. 
For each category, we sent a distinct query to GLOSS: each agent was instructed to \emph{``Generate a paragraph related to [category],''} alongside guidance drawn from our literature review:\emph{``retrospective summaries for RFMs should be high-level, qualitative, context-aware, customized, privacy-preserving, and designed to foster trust...''} 
GLOSS then determines which modalities, including additional VI outputs such as hourly summaries, are most relevant to each category and queries the corresponding database. 
The resulting category-level summaries are combined and reconstructed into a single daily summary. 
The system diagram is shown in Figure~\ref{fig:initial}, the prompts are attached in Supplementary Material.

To collect the underlying data used to generate the sample summaries, we relied on a multi-modal sensing infrastructure comprising demographic surveys (basic demographics, self-described living situation); smartphone sensing (GPS location mapped to self-reported, regularly visited location labels; screen lock/unlock status and brightness; Bluetooth and Wi-Fi connection status; Inertial Measurement Unit (IMU)-based activity labels from iOS); wearable devices (heart rate, respiration, IMU-based activity labels produced by an open-source human activity recognition model, e.g., sleep, sedentary, light, moderate, and vigorous~\cite{chan2024capture}, and step counts); and a conversational agent (conversation logs). 
Using this infrastructure, we collected over two months of real-world data from an older adult living alone, which we used to generate the retrospective summaries. 
We generated example summaries for four distinct days, incorporating both full-day and partial-day data to represent a range of realistic scenarios. The original data collection study was approved by our institution's Institutional Review Board (IRB).

Because our system largely adopts the summarization structure of VI and GLOSS, we expect a comparable level of factual accuracy (91.49\% and 87.93\% for the original VI and GLOSS systems, respectively). 
The initial summaries served as a probe to inspire participant feedback, not as a fully polished system. 
Accordingly, a researcher manually reviewed the four days of summaries generated for the interviews against the raw data using the VI dashboard, ensuring variety across days with differing data availability and checking each fact for accuracy; no objectively incorrect facts were found.

\subsection{Design of Interviews with RFMs and OAs}
We interviewed 12 RFMs and 2 OAs with diverse demographics and living situations (Table~\ref{tab:demographics}), recruited via social media ads and in-person outreach at a senior independent living community. Eligible participants (excluding those who lived together) completed a consent form and demographic survey before a one-hour Zoom interview. Recruitment for RFMs concluded upon reaching thematic saturation~\cite{Guest2006HowMI,Ando2014AchievingSI}; for OAs, we stopped early, because meaningful feedback on privacy would require real-world deployment (more described in Section 4.4.4).

Interviews began with questions on participants' caregiving situations, communication practices, and needs. Researchers then screen-shared sample summaries for participants to review and think aloud, followed by semi-structured questions on granularity, usefulness, format, trust, explainability, and privacy, and an invitation to imagine how such summaries might apply to their own family members. Throughout, the summaries served to elicit improvement ideas informing later deployment, not to evaluate performance. Interviews were recorded only with consent, anonymized via participant IDs, and compensated with a \$15 gift card; the study was approved by our institution's IRB.

Interview transcripts were analyzed using inductive thematic analysis \cite{corbin1990grounded,braun2006using,glaser1968discovery}: two HCI researchers with prior experience studying older adult populations independently coded the transcripts, discussed emerging themes, and iteratively merged codes by consensus into a shared codebook.
\begin{table}[t]
\centering
\small
\begin{tabular}{
  p{0.5cm}  
  p{0.5cm}  
  p{1 cm}  
  p{3.5 cm}  
  p{3.2cm}  
  p{4.5cm}  
}
\toprule
\toprule
\textbf{PID} & \textbf{Age} & \textbf{Gender} & \textbf{Race} & \textbf{Relationship to OAs/RFMs} & \textbf{Living Situation} \\
\midrule
C1  & 62 & Female & White & Child & Different cities or states \\
C2  & 50 & Female & White & Child & Same neighborhood \\
C3  & 65 & Female & White & Sibling & Same neighborhood \\
C4  & 41 & Female & White & Child & Different cities or states \\
C5  & 22 & Male & Black / African American & Child & Different cities or states \\
C6  & 61 & Female & White & Child & Same city/town \\
C7  & 27 & Female & Black / African American & Grandchild & Different cities or states \\
C8  & 36 & Female & White & Grandchild & Different cities or states \\
C9  & 65 & Female & White & Child & Different cities or states \\
C10 & 59 & Female & White & Child & Same neighborhood \\
C11 & 24 & Male & Black / African American & Child & Same city/town \\
C12 & 31 & Female & White & Child & Same city/town \\
P1 & 83 & Female & White & Grandchild & Same city/town \\
P2 & 71 & Female & White& Child & Same city/town \\
\bottomrule
\bottomrule
\end{tabular}
\caption{Demographic information of RFM and OA participants (PID: Participant ID).}
\label{tab:demographics}
\Description{Demographic table with 14 rows (12 remote family member participants labeled C1 through C12, and 2 older adult participants labeled P1 and P2) and 6 columns: Participant ID, Age, Gender, Race, Relationship to the older adult or family member, and Living Situation. Ages range from 22 to 83. Most participants are female (11 of 14) and White (10 of 14); the remaining participants are Black or African American (3 of 14). Relationships include child (9 participants), grandchild (3 participants), and sibling (1 participant). Living situations vary from living in different cities or states, to the same neighborhood, to the same city or town.}
\end{table}

\subsection{Results from the Interviews}
\subsubsection{RFM's Emotional Responsibility}

Throughout the interviews, all participants expressed positive feedback about the summary format, noting that this is something they have long wished for and that it would give them greater peace of mind.
C7, whose grandmother already uses an Apple Watch, reported that the summaries provided more information than she currently receives, making her feel that she \emph{``doesn’t really have to stress over trying to know why''} from massive data.
Regarding the granularity, the majority of RFMs expressed a desire for even more comprehensive information \emph{``as much as possible'' (C2)}. 
Unlike clinicians or professional caregivers, RFMs appreciate the level of detail since they carry a deep emotional responsibility to ensure their loved one is doing well because \emph{``It's your loved one, right? Not someone you just care for. (C5)''}.
This emotional connection further shaped preferences for tone of the summaries. RFMs felt that personalized, warmer language such as \emph{`your loved one'} instead of \emph{`user'} made the summaries more comforting and helped them feel their loved one was being cared for.

\subsubsection{From `What' to `How and Why': RFMs' Sensemaking Process}
Aligned with the strong sense of emotional responsibility, RFMs were cautious about the information presented and sometimes dissatisfied with overly aggregated statements: simply glancing at a summary and assuming everything is fine felt insufficient.
As a result, RFMs started to engage in their own sensemaking, especially when anomalies appeared: 
\begin{quote}
    \emph{Why was the Wi-Fi disconnected? Like, what really happened? [...] Is it, like, Grandma's fault or the device's fault? (C7) }
\end{quote}

Although the sample summaries were generated from another older adult's data, many RFMs immediately related the information to their own family members, unprompted, interpreting it through their extensive personal knowledge of their loved ones' routines. 
For instance, C8 noted that high morning activity in the sample would be unusual for their grandmother, who is typically active in the early afternoon rather than the morning; C10 similarly pointed to her mother's habitual morning coffee as context she would naturally apply when reading a summary.
While RFMs wanted summaries that incorporated this kind of personal context, most still expected to remain the primary interpreters of their loved one's life, drawing on their own knowledge to fill in gaps rather than relying solely on the system. 

In short, RFMs wanted summaries to go beyond a high-level account of \emph{what} their loved one was doing, toward scaffolding \emph{how} they were doing and \emph{why}, while still approaching the data cautiously: grounding their interpretation in trusted sources first, then layering on their own personal knowledge to make sense of it.

\subsubsection{Summary Content and Format}
RFMs also provided detailed feedback on both the content and format of the summary. 
The major categories, weather, device compliance, movement, crucial activities, and emotional wellbeing, remained important across participants. 
Beyond these categories, some RFMs also highlighted physiological health metrics, such as heart rate and respiration, as a distinct and essential category. 
For movement and crucial activities, RFMs’ sensemaking processes revealed that they were tightly interconnected around a whole picture of OA's day. 
Although RFMs broadly shared these categories of interest, specific needs varied by family: weather mattered most for RFMs living far away, especially in regions prone to extreme conditions like hurricanes, but less so for those living nearby; C8 wanted emphasis on when her loved one returned from church, C5 wanted exact heart rate values, and C1 wanted nap times.

Regarding the format of the summaries, the majority of RFMs liked the categorization of information, which made it easier to navigate and locate useful information by topic. 
Some participant, like C6, also noted that a simple, one-line summary capturing their loved one’s overall well-being would be especially helpful as a high-level overview on busy days.
Some participants, however, expressed a desire for paragraph-style summaries more like a story of their day. 
C1 and C7, for instance, felt that the current bullet-point format fragmented information and sometimes caused context to be lost. 
This preference was largely driven by the importance of maintaining a coherent narrative flow in the summary.


\subsubsection{Privacy and Autonomy}
Most RFMs reported that the older adults they cared for would be willing to share information like the summaries presented by us with family members: 
\begin{quote}
    \emph{``I don't think there's anything in here that would violate her privacy. I think she'd be, totally fine with this. It's just… she would feel like it was letting us know everything was going smoothly.''} (C10)
\end{quote}
We also conducted interviews with two older adults, focusing more closely on privacy and autonomy. 
Consistent with what RFMs expressed, the current version of the summaries did not appear intrusive to either participant. 
P1 commented that she was happy to share anything in the summaries with her granddaughter and could not think of anything she would want to restrict. 
P2 expressed a similar sentiment: 
\begin{quote}
    \emph{``I'm comfortable sharing anything like that because it doesn't bother me. I'm an open book. I mean, I\ldots I could tell her I have nothing to hide.''}   
\end{quote}

Through interviews with OAs, we found that showing participants sample summaries alone was unlikely to surface genuine insights around privacy concerns. 
At the same time, RFMs' needs were also highly personalized depending on different care situations. 
As a result, rather than pursuing more interviews, we moved forward with a real-world deployment to learn how these concerns play out in everyday, in-the-wild practice.

\section{Redesign of the Summary for Real-world Deployment}
\textbf{Weather}, \textbf{Device Usage}, and \textbf{Emotion} largely retained the structure of the initial summaries, aside from renaming ``Compliance'' to ``Usage'' for clarity. 
From the interviews, we found that activities and movement are highly interconnected, and that participants hoped to see a holistic picture of the OA's day, so we combined these two categories into a single \textbf{Activity} category. We also added a distinct \textbf{Health} category for physiological data, as requested by RFMs from the interviews.

To ensure easy navigation while sustaining narrative flow, we made several changes. First, we added a \textbf{One-sentence Summary} at the start of each report for a quick glance across all categories. For each category, we then layered the content into \textbf{Objective Statistics} (e.g., step counts under Activity, heart rate range under Health), followed by a more comprehensive, paragraph-style \textbf{Context-aware Description} for coherent narrative flow. These layers also aim to disclose information gradually, avoiding the overly aggregated presentation that interview participants found unsatisfying, and supporting RFMs' own sensemaking as they move from ``what's'' in the data to ``how'' their loved one is doing.
To account for RFMs' emotional responsibility, we made sure the summaries used phrases like ``your loved one'' instead of ``user'' to convey positive reassurance.

For real-world deployment, we added a \textbf{Real-time Data} section showing step counts from midnight to the present, offering a lightweight check-in option before the full summary is ready. 
Recognizing how OAs and RFMs might communicate around the summaries, we also added a \textbf{Next Time You Talk} section, generating discussion topics from the summary to help RFMs spark conversations with their loved ones.
For the deployment, we rendered each section/category in a fixed, consistent layout on the webpage.
The changes to categories were shown in Figure~\ref{fig:initial}, and sample summaries are shown in Figures~\ref{fig:checkin1}--\ref{fig:checkin5}.

\section{Case Study}
After finalizing the summary design for the deployment, we began recruiting and preparing to study how families share the summaries in their real lives.
\subsection{Study Design}
We recruited one OA–RFM pair from a collaborating senior independent living community in the southeastern United States, who expressed interest after we distributed a recruitment announcement through the community. 
This pair consists of a healthy, highly independent 81-year-old woman and her son, who lives only about 20 minutes away, with a weekly phone call and no substantial caregiving responsibilities (more details in Section 6.3). 
In other words, this is almost a least-likely case for dramatic tensions: the system was not being introduced into crisis care, dementia care, or a high-conflict family. 
We started with this relatively ``low-stakes'' dyad, hoping to surface potential tensions that could inform how such tensions might become more substantial in dyads with greater caregiving needs.

We customized the sensing setup for this family: because O1 does not own a smartphone, we installed the phone app on an iPad we provided and did not collect any phone sensing data, and used the iPad to connect the same smartwatch used in the previous data collection study. 
We installed an additional set of smart home sensors (floor plan attached in Supplementary Material), comprising 33 sensors across seven areas of the apartment (two bedrooms, two bathrooms, the foyer, kitchen, and living room), including 17 motion sensors (e.g., bedroom, bathroom, seating, closet), 12 contact sensors (e.g., doors, cabinets, drawers, refrigerator), and 4 power sensors (bedside lamp, microwave, stovetop, and TV).
Home sensor data was summarized similarly to other modalities, converted into sensor summaries such as: \emph{``The motion sensor in [AREA] was triggered at [TIME].''}~\cite{shi2026trace}.
Because O1 did not own or choose to have a conversational agent, we respected her preference and, unlike our previous data collection, did not have access to conversational logs.

\begin{figure}
    \centering
    \includegraphics[width=1\linewidth]{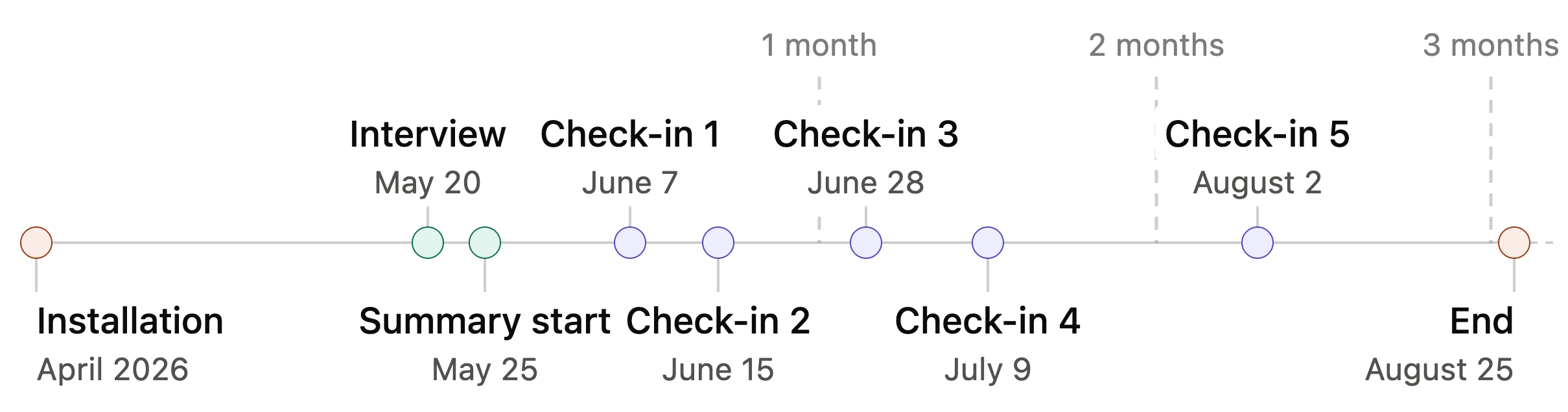}
    \caption{Timeline of device installation, participant interview, and follow-up check-in assessments from April through August 2026}
    \label{fig:timeline}
    \Description{Horizontal timeline from device installation to end of study, spanning April to August 2026. Nine points are marked in sequence: Installation (April 2026), Interview (May 20), Summary start (May 25), Check-in 1 (June 7), Check-in 2 (June 15), Check-in 3 (June 28), Check-in 4 (July 9), Check-in 5 (August 2), and End (August 25).}
\end{figure}

Sensors and devices were installed in April 2026 (Figure.~\ref{fig:timeline}). 
Following device setup, the first several weeks served as a technical validation period during which no summaries were delivered; data flow was monitored for reliability, and summaries were reviewed by researchers only. 
Before starting to send the summaries, we conducted separate interviews with O1 and F1 following the same interview protocol as Section 4. 
Based on this feedback, the summary format was refined, and daily summary sharing began for both participants several days after.

During deployment, each day's summary was generated one hour before the delivery time agreed upon by O1 and F1. 
A researcher reviewed each summary; if invalidity was identified (e.g., summary contains clearly irrelevant content), no summary was delivered that day. 
Upon passing review, the summary was released to O1 via email notification, accessible through a secure, password-protected website. 
O1 had approximately one hour to review the summary and optionally remove sentences or suppress the entire summary before it was forwarded to F1; if she did not respond within this window, the summary was sent to F1 by default. 
This default pipeline was discussed and agreed upon with both O1 and F1 at the outset. 
Since our interview studies did not surface much insight related to privacy, we designed this pipeline to give OAs simple, low-burden control over what content was sent, in hopes of surfacing more insight during the case study.
Both O1 and F1 could provide feedback on each summary directly on the same webpage through an open-ended text field, and past summaries remained accessible for review.

Once our system started delivering daily summaries, we conducted check-in interviews separately with O1 and F1, held roughly every 1–2 weeks depending on availability, to discuss their experiences; insights gathered were translated into prompts and pipeline adjustments to iteratively tailor the summaries throughout the deployment. 
After six weeks, once feedback stabilized, we moved to monthly check-ins and concluded data collection after 3 months. 
Participants received \$25/month for the sensing deployment and \$25/hour for interviews.
During the exit interview, the dyad expressed a wish to continue receiving the summary; we kept it enabled, but made it explicit that it was no longer monitored by the research team and that the study had ended.
This study was approved by our institution's IRB.

\subsubsection{Deployment of the Summary on the Webpage}
\begin{figure}
    \centering
    \includegraphics[width=1\linewidth]{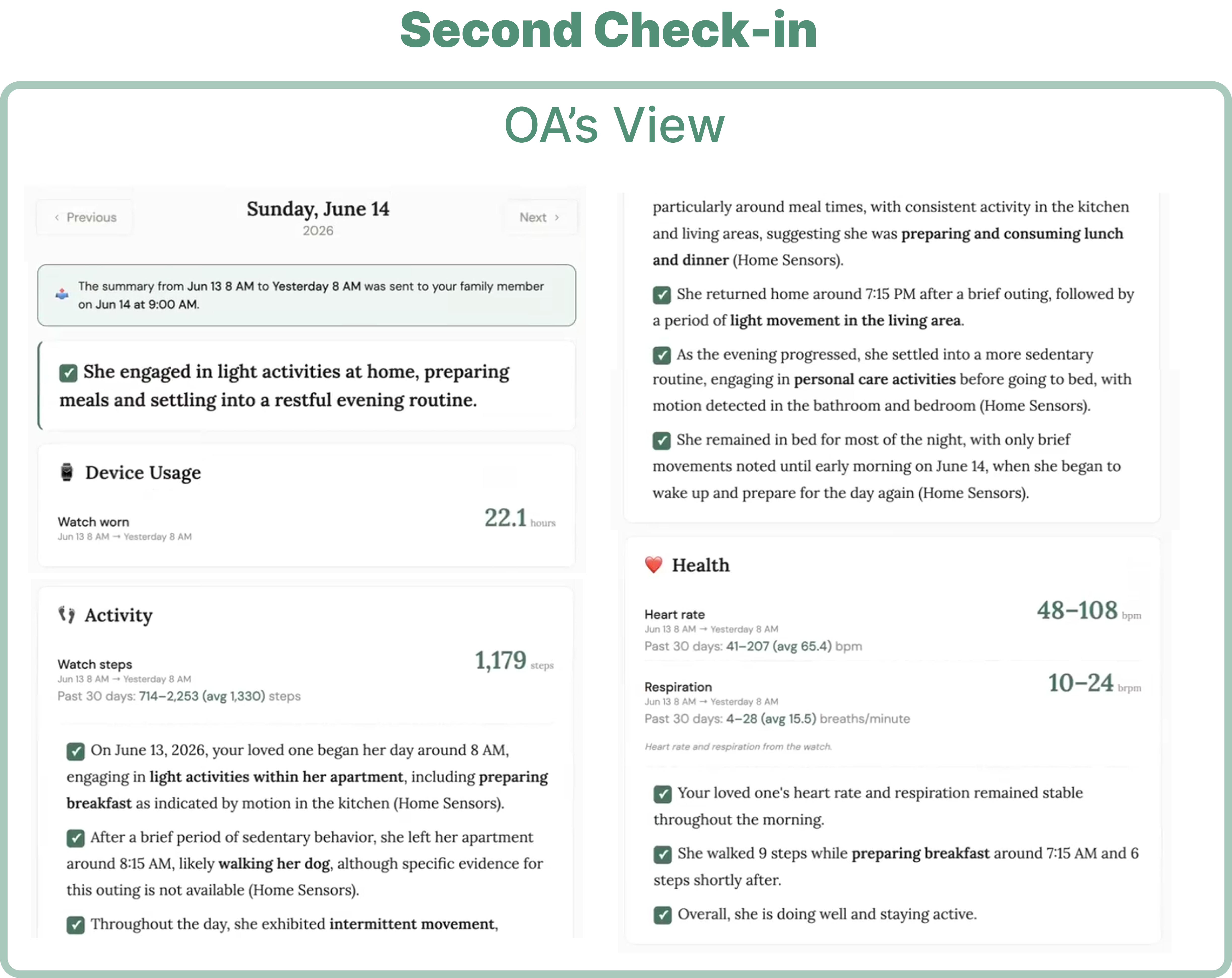}
    \caption{Summaries shown to O1 and F1 during second check-in.}
    \label{fig:checkin2}
    \Description{Screenshot titled 'Second Check-in,' OA's View, for Sunday, June 14, 2026. Notice: 'The summary from Jun 13 8 AM to Yesterday 8 AM was sent to your family member on Jun 14 at 9:00 AM.' One-sentence summary: 'She engaged in light activities at home, preparing meals and settling into a restful evening routine.' Device Usage: Watch worn 22.1 hours (Jun 13 8 AM → Yesterday 8 AM). Activity: Watch steps 1,179 (Jun 13 8 AM → Yesterday 8 AM; past 30 days 714–2,253, avg 1,330 steps), with narrative bullets: 'On June 13, 2026, your loved one began her day around 8 AM, engaging in light activities within her apartment, including preparing breakfast as indicated by motion in the kitchen (Home Sensors).' 'After a brief period of sedentary behavior, she left her apartment around 8:15 AM, likely walking her dog, although specific evidence for this outing is not available (Home Sensors).' 'Throughout the day, she exhibited intermittent movement, particularly around meal times, with consistent activity in the kitchen and living areas, suggesting she was preparing and consuming lunch and dinner (Home Sensors).' 'She returned home around 7:15 PM after a brief outing, followed by a period of light movement in the living area.' 'As the evening progressed, she settled into a more sedentary routine, engaging in personal care activities before going to bed, with motion detected in the bathroom and bedroom (Home Sensors).' 'She remained in bed for most of the night, with only brief movements noted until early morning on June 14, when she began to wake up and prepare for the day again (Home Sensors).' Health: Heart rate 48–108 bpm (past 30 days 41–207, avg 65.4 bpm); Respiration 10–24 brpm (past 30 days 4–28, avg 15.5 breaths/minute); note 'Heart rate and respiration from the watch.' Narrative bullets: 'Your loved one's heart rate and respiration remained stable throughout the morning.' 'She walked 9 steps while preparing breakfast around 7:15 AM and 6 steps shortly after.' 'Overall, she is doing well and staying active.'}
\end{figure}

\begin{figure}
    \centering
    \includegraphics[width=1\linewidth]{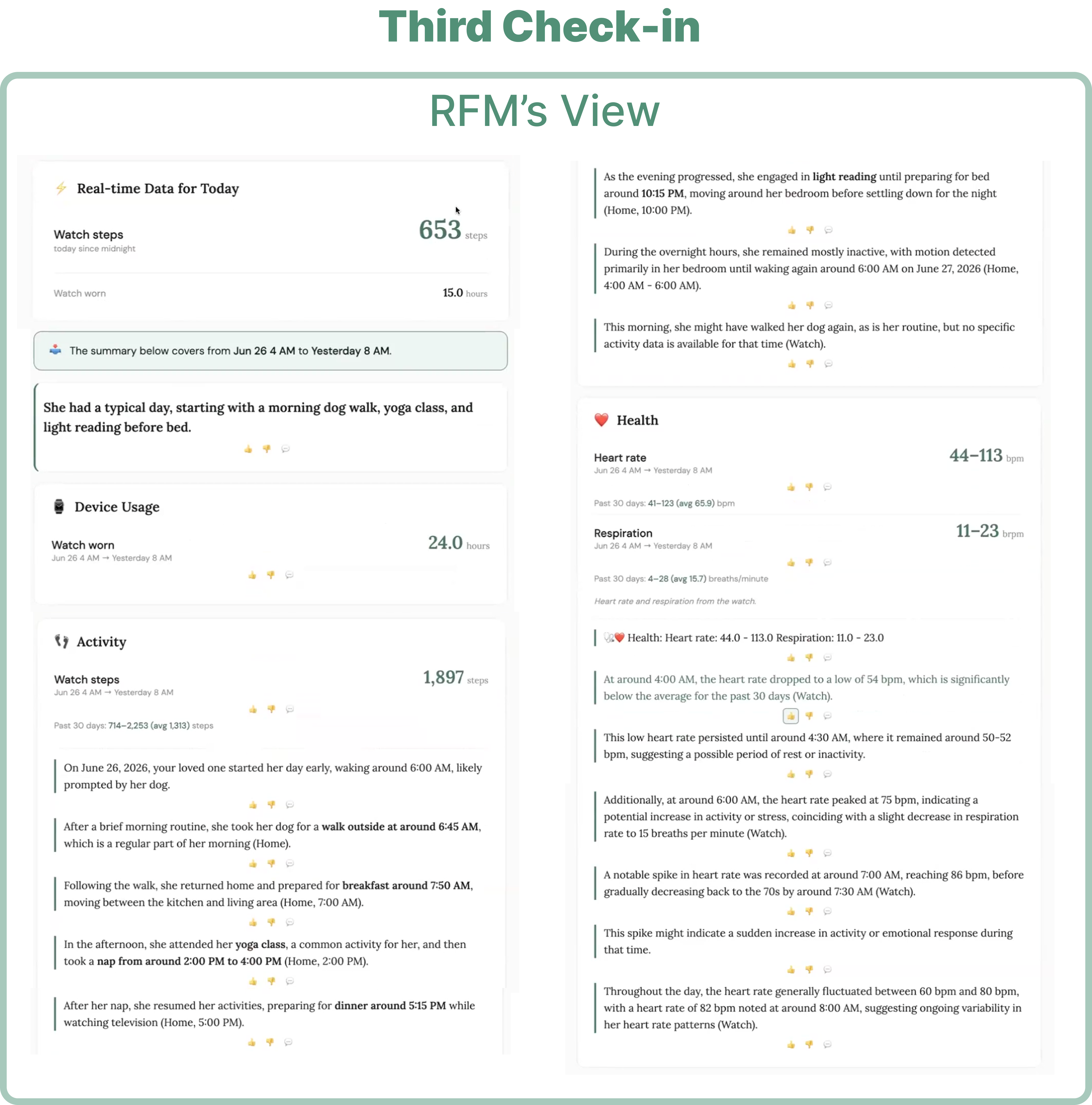}
    \caption{Summaries shown to O1 and F1 during third check-in.}
    \label{fig:checkin3}
    \Description{Screenshot titled 'Third Check-in,' RFM's View. Real-time Data for Today: Watch steps 653 (today since midnight); Watch worn 15.0 hours. Notice: 'The summary below covers from Jun 26 4 AM to Yesterday 8 AM.' One-sentence summary: 'She had a typical day, starting with a morning dog walk, yoga class, and light reading before bed.' (with feedback icons). Device Usage: Watch worn 24.0 hours (Jun 26 4 AM → Yesterday 8 AM). Activity: Watch steps 1,897 (past 30 days 714–2,253, avg 1,313 steps), with narrative bullets (each with feedback icons): 'On June 26, 2026, your loved one started her day early, waking around 6:00 AM, likely prompted by her dog.' 'After a brief morning routine, she took her dog for a walk outside at around 6:45 AM, which is a regular part of her morning (Home).' 'Following the walk, she returned home and prepared for breakfast around 7:50 AM, moving between the kitchen and living area (Home, 7:00 AM).' 'In the afternoon, she attended her yoga class, a common activity for her, and then took a nap from around 2:00 PM to 4:00 PM (Home, 2:00 PM).' 'After her nap, she resumed her activities, preparing for dinner around 5:15 PM while watching television (Home, 5:00 PM).' 'As the evening progressed, she engaged in light reading until preparing for bed around 10:15 PM, moving around her bedroom before settling down for the night (Home, 10:00 PM).' 'During the overnight hours, she remained mostly inactive, with motion detected primarily in her bedroom until waking again around 6:00 AM on June 27, 2026 (Home, 4:00 AM - 6:00 AM).' 'This morning, she might have walked her dog again, as is her routine, but no specific activity data is available for that time (Watch).' Health: Heart rate 44–113 bpm (past 30 days 41–123, avg 65.9 bpm); Respiration 11–23 brpm (past 30 days 4–28, avg 15.7 breaths/minute). Narrative bullets: 'Health: Heart rate: 44.0 - 113.0 Respiration: 11.0 - 23.0.' 'At around 4:00 AM, the heart rate dropped to a low of 54 bpm, which is significantly below the average for the past 30 days (Watch).' 'This low heart rate persisted until around 4:30 AM, where it remained around 50-52 bpm, suggesting a possible period of rest or inactivity.' 'Additionally, at around 6:00 AM, the heart rate peaked at 75 bpm, indicating a potential increase in activity or stress, coinciding with a slight decrease in respiration rate to 15 breaths per minute (Watch).' 'A notable spike in heart rate was recorded at around 7:00 AM, reaching 86 bpm, before gradually decreasing back to the 70s by around 7:30 AM (Watch).' 'This spike might indicate a sudden increase in activity or emotional response during that time.' 'Throughout the day, the heart rate generally fluctuated between 60 bpm and 80 bpm, with a heart rate of 82 bpm noted at around 8:00 AM, suggesting ongoing variability in her heart rate patterns (Watch).}
\end{figure}

\begin{figure}
    \centering
    \includegraphics[width=1\linewidth]{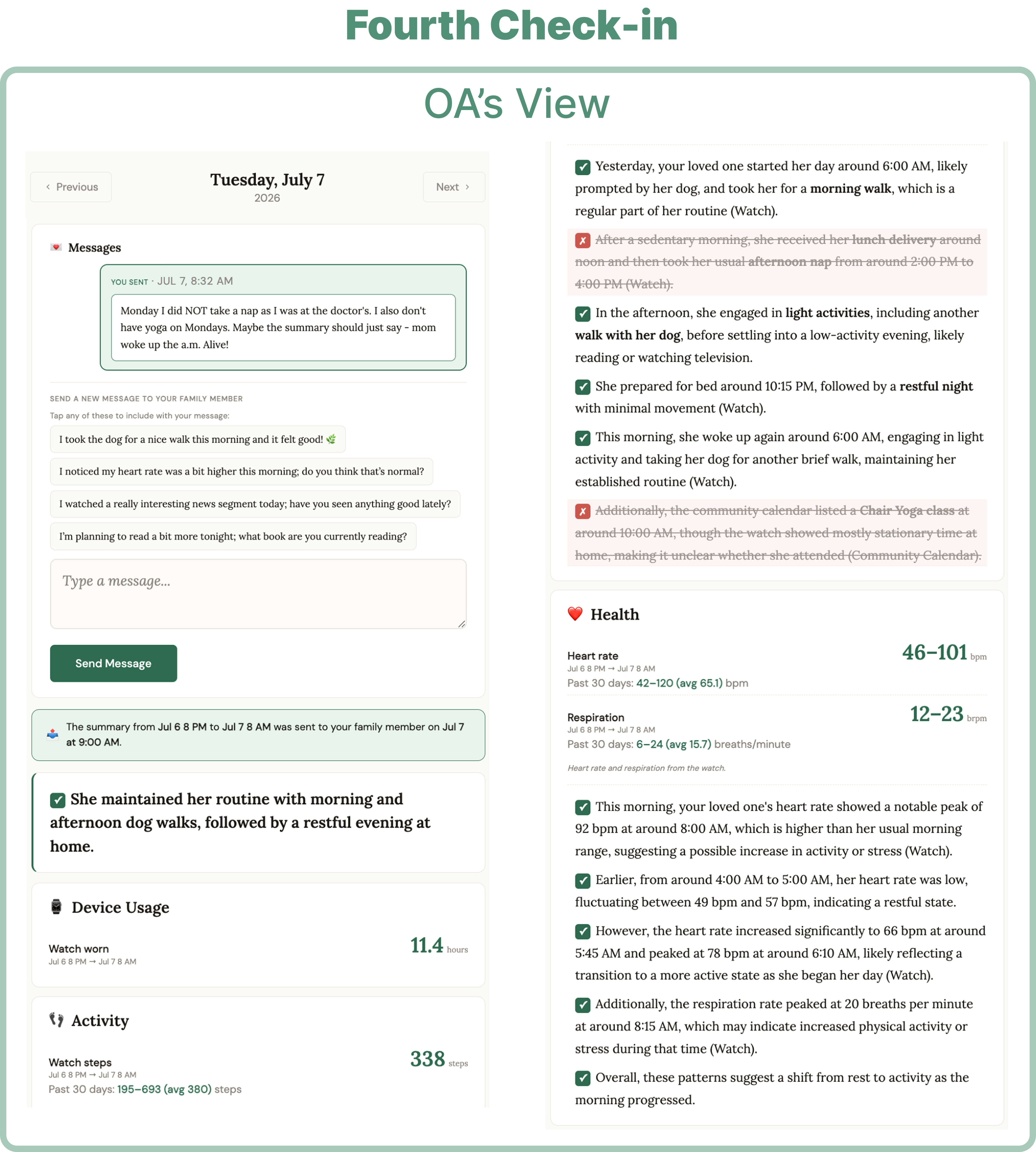}
    \caption{Summaries shown to O1 and F1 during fourth check-in.}
    \label{fig:checkin4}
    \Description{Screenshot titled 'Fourth Check-in,' OA's View, for Tuesday, July 7, 2026. Messages panel: a message the older adult sent on Jul 7 at 8:32 AM reading 'Monday I did NOT take a nap as I was at the doctor's. I also don't have yoga on Mondays. Maybe the summary should just say - mom woke up the a.m. Alive!' Below, four suggested reply prompts: 'I took the dog for a nice walk this morning and it felt good! ��', 'I noticed my heart rate was a bit higher this morning; do you think that's normal?', 'I watched a really interesting news segment today; have you seen anything good lately?', 'I'm planning to read a bit more tonight; what book are you currently reading?' with a message input box and Send Message button. Notice: 'The summary from Jul 6 8 PM to Jul 7 8 AM was sent to your family member on Jul 7 at 9:00 AM.' One-sentence summary: 'She maintained her routine with morning and afternoon dog walks, followed by a restful evening at home.' Device Usage: Watch worn 11.4 hours (Jul 6 8 PM → Jul 7 8 AM). Activity: Watch steps 338 (past 30 days 195–693, avg 380 steps), with narrative bullets: checked — 'Yesterday, your loved one started her day around 6:00 AM, likely prompted by her dog, and took her for a morning walk, which is a regular part of her routine (Watch).' crossed out/marked incorrect — 'After a sedentary morning, she received her lunch delivery around noon and then took her usual afternoon nap from around 2:00 PM to 4:00 PM (Watch).' checked — 'In the afternoon, she engaged in light activities, including another walk with her dog, before settling into a low-activity evening, likely reading or watching television.' checked — 'She prepared for bed around 10:15 PM, followed by a restful night with minimal movement (Watch).' checked — 'This morning, she woke up again around 6:00 AM, engaging in light activity and taking her dog for another brief walk, maintaining her established routine (Watch).' crossed out/marked incorrect — 'Additionally, the community calendar listed a Chair Yoga class at around 10:00 AM, though the watch showed mostly stationary time at home, making it unclear whether she attended (Community Calendar).' Health: Heart rate 46–101 bpm (past 30 days 42–120, avg 65.1 bpm); Respiration 12–23 brpm (past 30 days 6–24, avg 15.7 breaths/minute). Narrative bullets: 'This morning, your loved one's heart rate showed a notable peak of 92 bpm at around 8:00 AM, which is higher than her usual morning range, suggesting a possible increase in activity or stress (Watch).' 'Earlier, from around 4:00 AM to 5:00 AM, her heart rate was low, fluctuating between 49 bpm and 57 bpm, indicating a restful state.' 'However, the heart rate increased significantly to 66 bpm at around 5:45 AM and peaked at 78 bpm at around 6:10 AM, likely reflecting a transition to a more active state as she began her day (Watch).' 'Additionally, the respiration rate peaked at 20 breaths per minute at around 8:15 AM, which may indicate increased physical activity or stress during that time (Watch).' 'Overall, these patterns suggest a shift from rest to activity as the morning progressed.'}
\end{figure}

\begin{figure}
    \centering
    \includegraphics[width=1\linewidth]{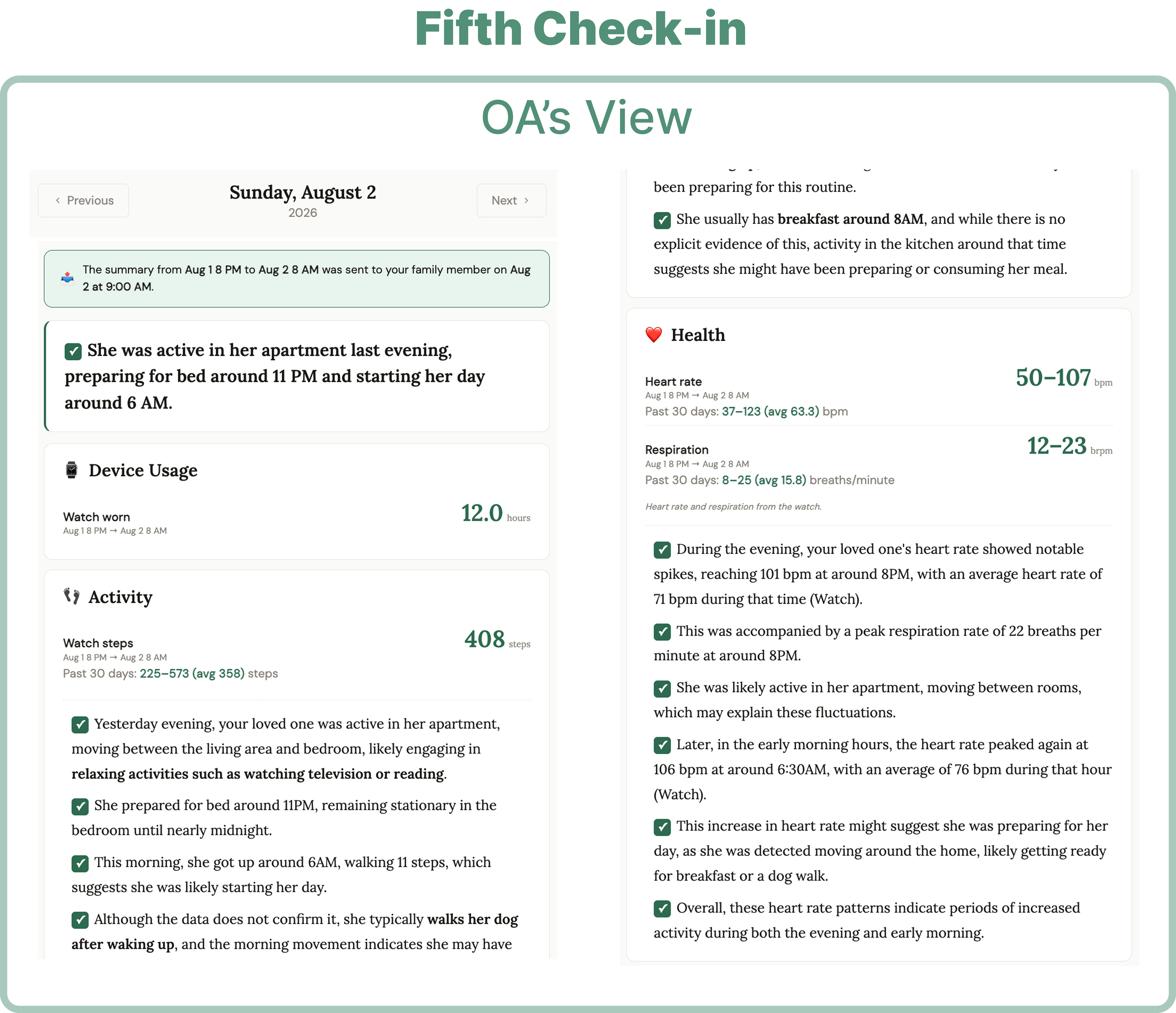}
    \caption{Summaries shown to O1 and F1 during fifth check-in.}
    \label{fig:checkin5}
    \Description{Screenshot titled 'Fifth Check-in,' OA's View, for Sunday, August 2, 2026. Notice: 'The summary from Aug 1 8 PM to Aug 2 8 AM was sent to your family member on Aug 2 at 9:00 AM.' One-sentence summary: 'She was active in her apartment last evening, preparing for bed around 11 PM and starting her day around 6 AM.' Device Usage: Watch worn 12.0 hours (Aug 1 8 PM → Aug 2 8 AM). Activity: Watch steps 408 (past 30 days 225–573, avg 358 steps), with narrative bullets: 'Yesterday evening, your loved one was active in her apartment, moving between the living area and bedroom, likely engaging in relaxing activities such as watching television or reading.' 'She prepared for bed around 11PM, remaining stationary in the bedroom until nearly midnight.' 'This morning, she got up around 6AM, walking 11 steps, which suggests she was likely starting her day.' 'Although the data does not confirm it, she typically walks her dog after waking up, and the morning movement indicates she may have been preparing for this routine.' 'She usually has breakfast around 8AM, and while there is no explicit evidence of this, activity in the kitchen around that time suggests she might have been preparing or consuming her meal.' Health: Heart rate 50–107 bpm (past 30 days 37–123, avg 63.3 bpm); Respiration 12–23 brpm (past 30 days 8–25, avg 15.8 breaths/minute). Narrative bullets: 'During the evening, your loved one's heart rate showed notable spikes, reaching 101 bpm at around 8PM, with an average heart rate of 71 bpm during that time (Watch).' 'This was accompanied by a peak respiration rate of 22 breaths per minute at around 8PM.' 'She was likely active in her apartment, moving between rooms, which may explain these fluctuations.' 'Later, in the early morning hours, the heart rate peaked again at 106 bpm at around 6:30AM, with an average of 76 bpm during that hour (Watch).' 'This increase in heart rate might suggest she was preparing for her day, as she was detected moving around the home, likely getting ready for breakfast or a dog walk.' 'Overall, these heart rate patterns indicate periods of increased activity during both the evening and early morning.}
\end{figure}

We use GPT-4o mini, a cost-efficient model, for summary generation throughout the longitudinal deployment.
No raw sensor data was passed to OpenAI; instead, queries were sent to GLOSS, which generated and ran code locally.
As shown in Figures \ref{fig:checkin1}--\ref{fig:checkin5}, the summary was displayed on a secure webpage hosted on an institution-owned server, with separate logins for O1 and F1 and a 10-minute idle auto-logout. 
The webpage also tracks user behaviors such as logins, activity, scrolling, and other interactions.
Before the summary was ready, the page displayed real-time data for both parties, shown in the top left of Figure~\ref{fig:checkin3}. 
Each summary began with a one-sentence overview, followed by sections on device usage, activity, and health.

Figure~\ref{fig:checkin1} shows the distinct views presented to O1 and F1. 
For O1, each AI-generated narrative sentence was marked with a green checkmark, and she could click on any sentence to suppress it (see the example in Figure~\ref{fig:checkin4}), preventing that content from being shared with F1. 
For F1, each sentence offered an option to rate it with a thumbs up or down or add a comment, entirely optionally (see the thumbs-up example in Figure~\ref{fig:checkin3}). 
Messages sent by O1 were displayed at the top of the page (see the example in Figure~\ref{fig:checkin1}, RFM view). 
Both O1 and F1 also had the option to send a message to the research team with additional feedback or comments at the end of the summary, and could click ``previous'' or ``next'' to navigate between different days' summaries.
\subsubsection{Data Analysis and Position}
One researcher served as the primary point of contact throughout the deployment to help build a trusted relationship with the dyad, while two additional researchers joined check-in interviews on a rotating basis.
We first analyzed engagement statistics by calculating login attempts, representing whether O1 and F1 checked the summary each day.
For qualitative analysis, we recorded each interview session, but rather than summarizing findings thematically, we report participants' insights and feedback chronologically, alongside our corresponding system iterations made in response to that feedback following RtD process as described in Section 3.1.
All three researchers, each with a background in human-AI interaction and prior experience working with older adults, collaboratively discussed each round of system iteration throughout the deployment and reached consensus on the final reporting of results presented in this paper.

\subsection{Current Caregiving Situation}
O1, an 81-year-old white woman, lives alone with her dog in a two-bedroom apartment in an independent living community with no professional caregivers. Her self-reported routine is highly regular: she wakes around 6AM, walks her dog before returning around 7AM, naps daily in the early afternoon, and goes to bed around 10:30PM. She receives one large midday meal from the facility, which she stretches to cover breakfast and dinner as well, and attends activities offered to community residents, such as exercise and yoga. She does not own a smartphone but is an active laptop user, checking email, browsing the internet, and following the news. She describes herself as highly independent.

F1, her 51-year-old son and only child, lives roughly a 20-minute drive (6–7 miles) away. He considers his mother healthy, active, and independent, with no significant caregiving needs or health monitoring currently in place, apart from a period of closer check-ins two years ago during her recovery from knee replacement surgery, when he temporarily moved in with her. Rather than providing hands-on daily care, he helps occasionally with errands, such as accompanying her to doctor's appointments, and with household tasks like moving or hanging items. Their communication is casual and low-key: a standing weekly Sunday-night phone call, supplemented by informal emails every few days, focused on everyday topics rather than health specifics. Despite living nearby, F1 visits infrequently due to his work schedule.


\subsection{Initial Needs for the Summary}
During the pre-deployment initial interview, both participants converged on wanting the summary to serve as light-touch reassurance rather than detailed monitoring. For F1, the core need was simple: knowing his mother was \emph{``up and alive,''} moving around, and had not fallen or had an accident. He wanted a low-effort way to check in without disrupting their casual communication style, and appreciated the concept of the daily summary we provided. O1 shared this same priority, wanting her son to know she had woken up safely each morning.
When reviewing O1's summary content from the past week, she was comfortable sharing all categories and content, but appreciated having the option to control what got sent. F1 described the summaries as \emph{``both interesting and useful,''} noting he especially liked the health section covering heart rate. 
Both O1 and F1 agreed the weather section was unnecessary, since F1 lives nearby and would have the same information anyway.
We also removed the Emotion section from the summaries, since no direct data (i.e., conversational logs) was available to inform its content.
Because both were primarily focused on knowing O1 was up and well in the morning, she requested that the summary's delivery time be moved earlier in the day, rather than sent as an end-of-day recap. After discussion, O1 and F1 were satisfied to begin receiving summaries in their current format and pipeline with some personalization: the weather and emotion sections were removed, and the summary would now be sent to F1 at 9AM, covering the window from 8AM the previous day to 8AM the current day.

\subsection{Compliance}
Over three months, we delivered the summary on 90 of 93 days; the three gap days were due to annual server maintenance (2 days) and system iteration (1 day). Researchers reviewed and approved all summaries through August 15 (80 days), 77 of them before 9 AM and the remaining 3 with some delay later the same day. After that point, by agreement with both parties, the study moved to a more casual cadence, and summaries were sent directly to O1 without approval.
O1 logged in on 86/90 days (95.6\%) and F1 on 81/90 (90.0\%), averaging 1.02 and 1.21 logins per active session, respectively. 
O1 suppressed a total of 10 sentences across 3 days, with no content deleted on any other day. 
O1 sent 5 messages to F1 and 3 to the research team; F1 sent none through the system.
Compliance for each check-in cycle is shown in Table.~\ref{tab:compliance}.

\begin{table}[t]
\centering
\small
\begin{tabular}{
  p{1.3cm}  
  p{0.9cm}  
  p{0.7cm}  
  p{0.7cm}  
  p{0.9cm}  
  p{0.9cm}  
  p{0.9cm}  
  p{0.8cm}  
  p{0.8cm}  
  p{0.8cm}  
}
\toprule
 & & \multicolumn{2}{c}{\textbf{Login}\textsuperscript{i}} & \multicolumn{2}{c}{\textbf{Visits/day}\textsuperscript{ii}} & & \multicolumn{3}{c}{\textbf{Message}} \\
\cmidrule(lr){3-4} \cmidrule(lr){5-6} \cmidrule(l){8-10}
\textbf{Period} & \textbf{Days} & \textbf{O1} & \textbf{F1} & \textbf{O1} & \textbf{F1} & \textbf{Suppr.}\textsuperscript{iii} & \textbf{O1$\to$F1} & \textbf{O1$\to$R} & \textbf{F1$\to$O1} \\
\midrule
Overall  & 90/93 & 95.6\% & 90.0\% & 1.02 & 1.21 & 10/3 d & 5 & 3 & 0 \\
\addlinespace
1st & 14/14 & 100\%  & 85.7\% & 1.21 & 1.33 & 0      & 2 & 1 & 0 \\
2nd & 8/8   & 100\%  & 100\%  & 1.12 & 1.00 & 0      & 0 & 0 & 0 \\
3rd & 11/13\textsuperscript{a} & 100\%  & 90.9\% & 1.00 & 1.30 & 0      & 0 & 0 & 0 \\
4th & 10/11\textsuperscript{c} & 100\%  & 100\%  & 1.00 & 1.18 & 2/1 d  & 1 & 0 & 0 \\
5th & 24/24 & 79.2\%\textsuperscript{b} & 91.7\% & 1.00 & 1.14 & 8/2 d  & 2 & 2 & 0 \\
Final & 23/23 & 100\%  & 78.3\% & 1.00 & 1.06 & 0      & 0 & 0 & 0 \\
\bottomrule
\end{tabular}
\caption{Compliance summary across the deployment. \textbf{Days}: days summary delivered/total days in period. \textsuperscript{i}\textbf{Login}: \% of days with at least one login. \textsuperscript{ii}\textbf{Visits/day}: average number of logins per day with at least one active login. \textsuperscript{iii}\textbf{Suppr.}: sentences suppressed/days with at least one suppression. \textbf{Message}: counts by direction (R = research team). \textsuperscript{a}2 days paused for server maintenance. \textsuperscript{b}O1 experienced an email delivery issue (went to spam folder). \textsuperscript{c}1 day paused for system iteration.}
\label{tab:compliance}
\Description{Compliance summary table with 7 rows, one for the overall three-month deployment and six for individual check-in periods (First through Fifth Check-in, plus a Final period), and 10 columns: Period, Days (summaries delivered out of total days), Login percentage for O1 and F1, average visits per active day for O1 and F1, sentences suppressed and days with suppression, and message counts by direction (O1 to F1, O1 to Researchers, F1 to O1). O1's login rate remained at or near 100 percent throughout the deployment except during the fifth period, when an email delivery issue reduced her login rate to 79.2 percent. F1's login rate ranged from 78.3 to 100 percent across periods. O1 suppressed content on 3 days total, concentrated in the Overall, Fourth, and Fifth periods; F1 never suppressed content. O1 initiated all messages sent through the system; F1 sent none.}
\end{table}

\begin{figure}
    \centering
    \includegraphics[width=1\linewidth]{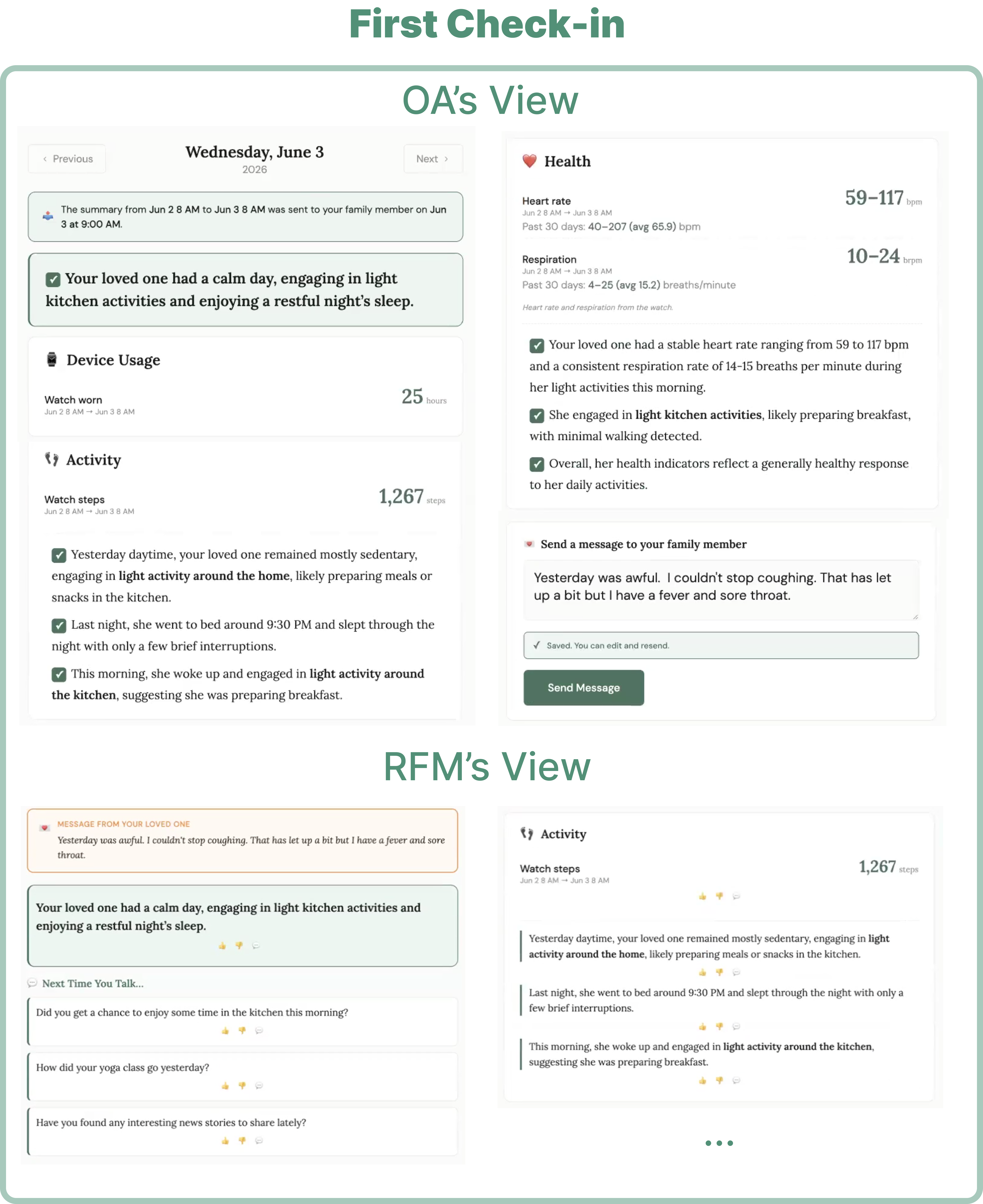}
    \caption{Summaries shown to O1 and F1 during first check-in.}
    \label{fig:checkin1}
    \Description{Screenshot titled 'First Check-in' showing two views for Wednesday, June 3, 2026: OA's View and RFM's View. OA's View, left panel: notice reading 'The summary from Jun 2 8 AM to Jun 3 8 AM was sent to your family member on Jun 3 at 9:00 AM.' One-sentence summary: 'Your loved one had a calm day, engaging in light kitchen activities and enjoying a restful night's sleep.' Device Usage: Watch worn 25 hours (Jun 2 8 AM → Jun 3 8 AM). Activity: Watch steps 1,267 (Jun 2 8 AM → Jun 3 8 AM), with narrative bullets: 'Yesterday daytime, your loved one remained mostly sedentary, engaging in light activity around the home, likely preparing meals or snacks in the kitchen.' 'Last night, she went to bed around 9:30 PM and slept through the night with only a few brief interruptions.' 'This morning, she woke up and engaged in light activity around the kitchen, suggesting she was preparing breakfast.' OA's View, right panel: Health: Heart rate 59–117 bpm (Jun 2 8 AM → Jun 3 8 AM; past 30 days 40–207, avg 65.9 bpm); Respiration 10–24 brpm (past 30 days 4–25, avg 15.2 breaths/minute); note 'Heart rate and respiration from the watch.' Narrative bullets: 'Your loved one had a stable heart rate ranging from 59 to 117 bpm and a consistent respiration rate of 14-15 breaths per minute during her light activities this morning.' 'She engaged in light kitchen activities, likely preparing breakfast, with minimal walking detected.' 'Overall, her health indicators reflect a generally healthy response to her daily activities.' Below, a 'Send a message to your family member' box contains the older adult's typed message: 'Yesterday was awful. I couldn't stop coughing. That has let up a bit but I have a fever and sore throat.' with a note 'Saved. You can edit and resend.' and a Send Message button. RFM's View, left panel: a message box labeled 'MESSAGE FROM YOUR LOVED ONE' repeating the same message about coughing, fever, and sore throat. Below it, the same one-sentence summary ('Your loved one had a calm day...') with thumbs-up/thumbs-down/comment icons. A 'Next Time You Talk...' section lists three suggested questions: 'Did you get a chance to enjoy some time in the kitchen this morning?', 'How did your yoga class go yesterday?', 'Have you found any interesting news stories to share lately?' each with feedback icons. RFM's View, right panel: Activity: Watch steps 1,267 (Jun 2 8 AM → Jun 3 8 AM), with the same three narrative bullets as OA's view, each accompanied by thumbs-up/thumbs-down/comment icons.}
\end{figure}
\subsection{First Check-in}
The first check-in was postponed for a week since O1 was ill.
 


\subsubsection{Feedback from O1 and F1}
O1 initially expressed dissatisfaction with the summary's accuracy, but when we reviewed several days together, we found few actual errors once she reflected on them. 
The only factual error she confirmed occurred on June 2nd, under the activity narrative: \emph{``Last night, she went to sleep around 11PM, with a brief interruption at around 5:30AM before waking up around 8AM (Watch).''} 
O1 noted that she had not gone back to sleep after 5:30AM. 
Aside from this, no factual errors were identified, a finding that surprised O1 herself, given how inaccurate she had initially perceived the summary to be: \emph{``Oh, actually, actually, that is accurate. (after reflecting and describing her whole day on June 3rd)''}

Upon further investigation, her perception of the summaries' inaccuracy is better understood as a sense of misinterpretation: although the facts were correct, the phrasing of the summaries gave her the impression that she \emph{``did nothing''} during the day.
\begin{quote}
\emph{``I don't really care, because I know what I'm doing, but the summary makes it look like I do absolutely nothing. I mean, you know, it says, oh, she had a restful night of sleep, she went to the, you know, used her microwave. And then… And then that's all I do. For instance, your activity trackers you know. they don't say that I leave the apartment, which… I do, I mean, I don't spend my life in the apartment. I mean, it's kind of a weird summary of… I mean, it's just not what happens during the day.''}
\end{quote}
This is the summary from June 3rd (also in Figure~\ref{fig:checkin1}):
 
\begin{tcolorbox}[title=Example Summary on June 3rd, colback=summary!5, colframe=summary!60, fontupper=\small]
Yesterday daytime, your loved one remained mostly sedentary, engaging in light activity around the home, likely preparing meals or snacks in the kitchen.
 
Last night, she went to bed around 9:30 PM and slept through the night with only a few brief interruptions.
 
This morning, she woke up and engaged in light activity around the kitchen, suggesting she was preparing breakfast.
\end{tcolorbox}

Although O1 confirmed there were no factual errors, she felt the description made it seem as though she had stayed home all day, portraying her as \emph{``inactive''}, a characterization she disagreed with, given that she knew she had walked the dog and engaged in some activities outside the apartment. She reasoned this might have stemmed from all sensors being installed indoors, leaving no way to capture her outdoor movement.


A similar discrepancy arose on a day O1 was ill: she found the summary's description of \emph{``restful''} inaccurate given how \emph{``miserable''} she had actually felt, but was more forgiving here, framing it as a sensing limitation rather than a system failure:
\begin{quote}
\emph{``I probably sat down a lot, and the sensors probably just thought I was having a great old time. But that's not the sensor's fault, right?''}
\end{quote}
She chose not to remove this content before it was shared with F1 (\emph{``No, I mean, why would I bother.''}), instead sending him a direct message:
\begin{tcolorbox}[title=Message sent from O1 to F1 on June 3rd, colback=message!5, colframe=message!60, fontupper=\small]
Yesterday was awful. I couldn't stop coughing. That has let up a bit but I have a fever and sore throat.
\end{tcolorbox}

F1 similarly acknowledged the sensors' limits while valuing the message:
\begin{quote}
\emph{``I understand that your sensors can't do everything, it'd be great to somehow monitor that she was sick, but I don't… I mean, that gets pretty intense and invasive. So, I guess that's why we have the note.''}
\end{quote}

Asked about her expectations after sending this message, O1's response was direct: \emph{``Nothing.''} 
She explained that her son had a busy work schedule, spanning early morning to late night, so she did not anticipate a response: \emph{``I wasn't waiting for him to react.''} 
She would, however, expect a response in a more severe situation, such as hospitalization, and would reach out directly by phone or email: \emph{``I wouldn't do it through this (summary system).''} She did, however, still expect F1 to \emph{``glance at it every day to make sure I woke up,''}.

A second major theme, emphasized more by F1, concerned the summaries' repetitiveness:
\begin{quote}
\emph{``I became used to it. I'm like, oh, this is just the same thing every day.''}
\end{quote}
He attributed this to the sensing system's constraints, and that his mother's daily routine is itself repetitive. 
The repetitiveness did not undermine his trust, though he remained, in his own words, \emph{``greedy''} for more information.
While describing this desire for detail, F1 suddenly realized the lack of detail might instead stem from his mother suppressing content, something he would have no way of knowing, rather than from the summary itself. He clarified he would be completely fine with that.

Most changes we applied to the redesigned summaries were well received. 
F1 liked the one-sentence summary and Activity section for giving him an overall picture of O1's day. He also cared about the Health section, both the objective statistics and the context-aware description, and reported observing that O1's heart rate seemed to elevate on later days of the week, planning to monitor this further.
Though not through the system, both O1 and F1 mentioned discussing several unusually low respiration readings in the summary. O1, however, reported that she had not even noticed these readings; it was F1 who brought them to her attention. 
F1 also found the underlying data sources (e.g., Watch, Home Sensor) useful, and valued affirming content such as \emph{``She appeared to be in good health, there's no significant concerns noted''} as reassurance.
F1, however, showed little interest in the ``Next Time You Talk'' feature (\emph{``I don't have a problem striking up a conversation with my mother...''}), and O1 was similarly hesitant to adopt proposed changes to the message panel, such as adding emojis (\emph{``I don't think… I don't know that his brain works that way.''}).

Given the high engagement, we also asked both participants about their motivation for checking the summary nearly every day. F1 explained he did not want to miss a day, both for his mother's safety and to give the researchers useful feedback; multiple same-day logins, he said, often reflected forgetting whether he had already checked that morning. O1's motivation was more idiosyncratic: she enjoyed identifying \emph{``what kind of mistakes it made,''} and noted that checking each morning had become almost routine.



\subsubsection{System Iteration}
Delving deeper into the inaccurate portrayal, we traced the root cause to a lack of tracking outside the apartment, since the home sensors captured only indoor movement. 
Without explicit instructions addressing this, the system disproportionately emphasized activities with more available data, such as movement in the kitchen, captured by multiple motion and contact sensors placed throughout the home. 
When O1 was outdoors, by comparison, the only available data was step counts from the watch (which also tracked steps indoors, making outdoor steps less distinguishable) and the front foyer sensor, so outdoor activity was often underrepresented and overshadowed by denser kitchen-related readings. 
To address this, we modified the prompt to rebalance the summary, adding a line to emphasize leaving and returning to the apartment via the foyer sensor: \emph{``Always include leave/enter apartment and door activities if given, or infer from sensors close to the front door.''}
 
Although the LLM had access to participant demographics, it lacked context for interpreting metrics like step counts, leading to potentially biased conclusions such as \emph{``minimal movement.''} We addressed this by adding a Longitudinal Context section to the GLOSS queries, including 30-day statistics (average, range) for health (heart rate, respiration) and activity (step count) metrics, provided both to the model and displayed on the webpage beneath each daily value, informing the LLM's reasoning while giving F1 a baseline for his own judgment.
 
To address repetitiveness, we acquired a shared monthly community calendar from the independent living facility and incorporated it into the summary. Because F1 wanted more information, we also relaxed the earlier constraints limiting the summary to 3 sentences.
 
For the message panel, we recognized the system allowed only O1 to send messages to F1, with no way for him to respond, creating an imbalanced power dynamic. We addressed this by adding a mutual messaging feature (examples in Figure~\ref{fig:checkin4}), replacing the earlier ``Next Time You Talk'' section F1 had found unhelpful.
 
Implementing these changes was a rapid iteration process: we implemented these changes within two days of the interview, aiming to continue optimizing based on feedback in the following week.
\subsection{Second Check-in}
\subsubsection{Feedback from O1 and F1}
O1 confirmed the new summaries still had not captured her whole day, but noted, \emph{``it's getting a little bit better,''} and that many leave-apartment events were now captured. F1 agreed this was very useful:
\begin{quote}
\emph{``When she leaves the house, that's a good sign for me... leaving the home's more interesting to me. Her hanging out at the house doesn't really tell me much, you know?''}
\end{quote}
As before, we found no factual errors across reviewed days. 
Some events, however, were still missing. On Friday, O1 noted:
\begin{quote}
\emph{``Well, shortly before noon, I always leave my apartment, because I always walk my dog at noon, and none of them seem to be picking up on that.''}
\end{quote}
O1 continued to enjoy catching errors and became increasingly specific about expected content based on what had been captured correctly on other days. We probed further into her routines and identified several recurring patterns, e.g., that she walks her dog three times a day nearly every day; having seen the summary capture one outing, she now expected it to capture all three. 
She also noted she was typically active before 8AM, but since the summary's window only ran from the previous day's 8AM to the current day's, these morning activities went uncaptured.

F1 found the new summaries more engaging, noting differences in phrasing (e.g., variations in how nap time was described) and richer detail overall; he was comfortable with the resulting longer narrative. 
With this added detail, however, O1 began considering that some information might not be relevant to F1. She was initially dismissive about sharing community-calendar activities:
\begin{quote}
\emph{``Oh, I don't think that's important, because… I mean, he could care less, and I know what I'm doing.''}
\end{quote}
but F1 found it quite interesting and informative. Notably, O1 raised no privacy concerns nor asked to delete any details, her comment reflected only a suspicion that F1 might not need certain information, not a desire to withhold it.

F1 continued closely tracking the health metrics for signs of concern, finding none (\emph{``It does look like it's slightly ticking up, but not to a point where it would be concerning.''}). Though comfortable interpreting the numbers himself, he began to wish the health narrative offered more context for abnormal data, to add value beyond the numbers.

Neither party sent a message that week, O1 saw nothing noteworthy to report, and F1 remained reluctant to change his communication habits or use the system's messaging feature.


\subsubsection{System Iteration}
A major need was for the system to better integrate O1's routine, given how fixed it is. 
We incorporated the newly captured routine details into GLOSS's queries, with instructions emphasizing accurate reflection of this information: \emph{``- ROUTINE EXTRACTION \& MAPPING. Map routine activities {routine} to the most plausible sensor signature.''} 
To better capture her active mornings, we also changed the summary's data window to start at 4AM rather than 8AM the previous day, and added instructions directing GLOSS to focus on and explain health anomalies: \emph{``Summarize any notable patterns or anomalies, such as abrupt changes, unusually high or low values, or patterns tied to specific activities or times of day. Expand on the paragraph by adding clear explanations and context from all available sources.''}

\subsection{Third Check-in}

\subsubsection{Feedback from O1 and F1}
O1 immediately noticed the change in how health information was narrated and was not fully satisfied, feeling it overemphasized her heart rate fluctuations:
\begin{quote}
\emph{``There seems to be an emphasis on when my heartbeat goes up, but… You know, that's, like, just for probably a few minutes at the most. So it's kind of like… I don't… I don't want to alarm [Name of F1], because most of the time it's, you know, like in the 60s or 70s.''}
 
\emph{``[...]''}
 
\emph{``In fact, I went to the doctor this week, and my blood pressure was excellent. I don't want him to be concerned because there's sort of an emphasis on, Oh, your heart went up to, you know, this or that. You must have gotten excited or stressed or busy or something.''}
\end{quote}
She was also uncertain of the purpose of sharing these details: \emph{``I mean, what's he supposed to with the information. I don't know. I don't see the point of it.''}

F1, by contrast, consistent with the interest in more detail he had expressed the previous week, welcomed the new information:
\begin{quote}
\emph{``I think the biggest. Well, the only thing I really noticed was... I assume it was AI started like summarizing, or pointing out if there was a change in, like, heart rate or breath rate. I noticed that portion was different. That was interesting to me.''}
\end{quote}
He specifically appreciated the system proactively surfacing anomalies:
\begin{quote}
\emph{``It was just nice that there was a bullet point that said, hey, the heart rate was different, that was a little bit different than before, that it did that reasoning for me, and I didn't have to notice it on myself, because I probably wouldn't have caught it.''}
\end{quote}
One notable anomaly occurred on June 28th, for which F1 gave a thumbs-up:
\begin{tcolorbox}[title=Example Summary on June 28th, colback=summary!5, colframe=summary!60, fontupper=\small]
At around 4:00 AM, the heart rate dropped to a low of 54 bpm, which is significantly below the average for the past 30 days.
\end{tcolorbox}
He found this \emph{``a good use for AI''} but saw no need for action, relying on his own common sense: \emph{``Nothing alarming that it felt like she was sleeping, and her heart rate would be lower.''} 
He had not discussed these health-metric changes with O1, since nothing had stood out, though he would raise a clear trend directly if one emerged.

Overall, F1 remained noticeably more relaxed about the anomalies than O1, who also confirmed she was not uncomfortable sharing the information but was less certain how F1 would interpret or react to it, worrying it might introduce unnecessary concern on his part. 
Both noted the past two weeks had been especially busy for F1: they had canceled their usual Sunday call and had not spoken by phone in about two weeks, and so had had no chance to discuss the changes, further confirming O1 and F1's generally low-key communication style.

\subsubsection{System Iteration}
Although O1 did not ask us to remove any of the new health narratives, we took her perspective seriously and modified the prompt to soften how anomalies were described: \emph{``- Do not overinterpret standout patterns. Report them with context and potential reasons, but avoid drawing conclusions unless the values are clearly outside normal range. For example, a rising heart rate is worth noting with its likely cause, but if it remains within a normal range, do not flag it as a concern.''}

\subsection{Fourth Check-in}
\subsubsection{Feedback from O1 and F1}
O1 expressed frustration over a noteworthy error she noticed in this week's summary: rather than napping, she had been at a doctor's appointment. There was also a mismatch involving her yoga schedule:
\begin{tcolorbox}[title=Message sent from O1 to F1 on July 7th, colback=message!5, colframe=message!60, fontupper=\small]
Monday I did NOT take a nap as I was at the doctor's. I also don't have yoga on Mondays. Maybe the summary should just say - mom woke up the a.m. Alive!
\end{tcolorbox}
The correct yoga class was indeed listed on the community calendar, but O1 confirmed she had not attended it; the napping error likely arose from motion recorded around that time, with the LLM hallucinating a nap based on O1's regular schedule. 
She reported that she had also emailed F1 directly about the error.

Compared with her earlier excitement about catching errors, O1 now expressed growing frustration, her sense of the summary's value having shifted toward its role in sharing information with her son, making incorrect content feel more annoying, though she agreed the inaccuracy did not meaningfully affect anything: 
\begin{quote}
\emph{``I guess I just need to accept the inaccuracies. I mean, they don't really affect me, except that I see that they're inaccurate, which is kind of… I don't know why it annoys me, but it does. I mean, it's not a big deal, I know that.''}
\end{quote}
As she increasingly viewed the summary as meant for F1, she grew more curious about how he actually experienced it: 
\begin{quote}
\emph{``I'm just giving him my opinion. I don't know what he gets out of it. He's never really told me if he gets anything out of it. You guys probably have a better idea than I do, yeah.''}
\end{quote}

F1, by contrast, offered a much more relaxed take, with little additional feedback (\emph{``I'm pretty used to it at this point. I know what to expect. It's good.''}) and was unbothered by O1's message:
\begin{quote}
\emph{``I think the last line was just sort of a joke on her behalf. Like, I don't think she meant that in any kind of negative way. She was just kidding, is how I viewed that last one. And I think she was just clarifying… Yeah, she was just joking. She was just being silly. And then I think she was just clarifying that she didn't take a nap, which, you know… She went out to the doctors, so I think she was just giving feedback is how I took that.''}
\end{quote}
O1 did not ask us to change anything about the summaries in response to her comments.

We also asked O1 about the suggested-topics feature in the message panel, adapted from the earlier ``Next Time You Talk'' section; although she had sent several messages to F1, she had never used these prompts, expressing clear reluctance:
\begin{quote}
\emph{``I don't need prompts like that... I mean, I just wouldn't use that. I don't need, I mean, I can't imagine sending these to [Name of F1]. He'd think I lost my mind. Well, it doesn't sound like me, and he wouldn't be interested, I mean I can't imagine I would ever use these.''}
\end{quote}

\subsubsection{System Iteration}
Our earlier emphasis on capturing routines (Section 6.7.3) had inadvertently misled the system into over-relying on them, in this case causing it to report an afternoon nap that had not occurred and was not supported by sensor data. We therefore refined the prompt to keep mentioning O1's regular routines, but to explicitly indicate whether each was confirmed by the sensing data:
 
\emph{``- ROUTINE EXTRACTION \& MAPPING. Map routine activities {routine} to the most plausible sensor signature
    - Plausible match found  → name the routine with supporting sensor evidence.
    - Match is ambiguous     → name the routine with light hedging.
    - Match is missing or contradicting → still name the routine, with explicit hedging that
      cites the specific missing/contradicting sensor evidence.
      ''}
      
\subsection{Fifth Check-in}

\subsubsection{Feedback from O1 and F1}
We gradually settled into a more casual monthly check-in schedule rather than weekly, while still maintaining email contact with both parties. 
O1 reviewed her summary almost daily (aside from several times when the emails went to the spam folder) and had grown accustomed to reporting errors to both researchers and her son. 
On July 11th, right after the check-in interview, she told researchers: \emph{``The activity data seems more accurate.''} 
That same day, she suppressed the following line before it reached F1: \emph{``This spike is higher than the typical range for this time of day, suggesting possible light activity as motion was detected in the bedroom and kitchen during this period,''} consistent with her not wanting F1 to worry about her. 
On another day, she messaged researchers directly, rather than F1, to help align the data: \emph{``I did not wake up at 4 a.m. I suspect the dog got off my bed after cowering next to me due to storm.''} 
She neither sent this to F1 nor suppressed the corresponding sentence. Separately, she messaged F1 directly:
\begin{tcolorbox}[title=Message sent from O1 to F1, colback=message!5, colframe=message!60, fontupper=\small]
I am alive and well. Actually took a longer walk at 6:45. My watch says 530 right now (7:30 a.m.)
\end{tcolorbox}

O1 confirmed there was really no information she was absolutely unwilling to share, though she made more edits on days she had more time. 
She maintained her heart was fine and did not want to become, in her words, \emph{``crazy on her own heart rate''}: \emph{``I don't necessarily wanna… be that aware… of what my heart is doing. As long as it keeps beating.''} She did, however, recognize her son genuinely cared about this information, and was comfortable sharing it on that basis: 
\begin{quote}
    \emph{``Well, it depends on what he… what he's interested in.''}
\end{quote}

Revisiting the earlier message she had sent (\emph{``Maybe the summary should just say --- mom woke up the a.m. Alive!''}), O1 remained satisfied with the current summary and did not request changes, but we respected her sentiment and brainstormed adjustments with her via email. 
We ultimately shifted the summary's start window from 4AM to 8PM the previous evening, so it would focus on the previous night and current morning rather than the entire prior day. She did not feel this made a substantial difference and preferred we keep the new window going forward.

F1 remained satisfied with the summary, was comfortable with the shifted window, and had grown accustomed to the AI's phrasing style: 
\begin{quote}
    \emph{``I feel like the AI phrases it in such a way that it could be alarming, but I'm kind of used to the way it talks now, or… You know? So I'm like, oh, okay. And I'm, you know, aware of the numbers. I'm like, that's not nothing crazy... I think it kind of alarms you, and then it tells you what, you know, it's not really a big deal.''} 
\end{quote}

Having reached the default two-month participation period, both O1 and F1 expressed a desire to continue in the study.

\subsubsection{System Iteration}
During the final deployment period, we adjusted the summary generation window to span from 4AM the previous day to 8PM the current day, with no other changes made.


\subsection{Summary of Insights}
Throughout the case study, both parties actively engaged with the summaries, sustaining an engagement rate above 90\% over three months, suggesting the summaries successfully integrated into their existing routines and that insights from our initial interview study carried over meaningfully into deployment. 
At the same time, both participants' relationships to the summary shifted over this period: O1 moved from initial curiosity and excitement about catching errors, to a growing awareness that the summary was ultimately meant for F1, and a resulting annoyance at those same errors; F1 moved from active interest, to a sense of repetitiveness and a desire for more information, and gradually gained a better understanding of the AI's narrative style (e.g., how it described and explained heart rate patterns).
 
The case study, however, surfaced real-world tensions not observed in interviews with broader OAs/RFMs, or even the dyad themselves (e.g., O1 raised no concerns about sharing summaries with F1 when initially shown sample summaries). These insights fall into two categories: challenges in accurately representing the OA's day, where factual accuracy alone proves insufficient (Contribution 1), and tensions in how summaries were communicated and interpreted within the dyad, driven by uncertainty about how shared information would be perceived (Contribution 2). We unpack both in Sections 7.1–7.2, and present design implications.

\section{Discussion}
In this section, we situate our findings within existing literature, and report limitations of the study.


\subsection{Living with Imperfect Sensing in AI-Mediated Aging-in-Place Communication}

Prior work on sensing-to-summary systems has largely focused on improving interpretability at the model level: transforming raw sensor data into narrative representations \cite{li2025vital,englhardt2024classification,ouyang2024llmsense,fang2024physiollm}, then refining that interpretation through prompt engineering \cite{ji2024hargpt,cosentino2024towards,choube2025gloss}, role-play \cite{yang2024drhouse,ji2024hargpt}, or fine-tuning \cite{kim2024health,cosentino2024towards}. 
Our findings suggest this alone is insufficient: even a technically accurate summary can misrepresent someone's day, since accuracy involves not just \textbf{what} happened, but \textbf{how} someone's day felt. 
This limitation matters even as sensor-language models such as SensorLLM and SensorLM improve alignment between sensor signals and natural language \cite{li2025sensorllm,zhang2026sensorlm}: such advances may make interpretations more accessible and expressive, but also make incomplete evidence appear more coherent and authoritative without ensuring a faithful representation of lived experience.

The gap between ``what'' and ``how'' has multiple causes. One is prioritizing relevant information within large volumes of heterogeneous sensing data.
Although systems such as GLOSS address this with multi-agent architectures that use a coordinating agent to identify relevant modalities and perform targeted function calls \cite{choube2025gloss,le2025multi,yoon2026consensus}, with only general prompt instructions, the volume of indoor motion-sensor data still biased our summaries toward over-representing indoor activity. 
Directing more targeted prompts to GLOSS (e.g., prioritizing front-door foyer data for leave/enter-home activity) rebalanced this emphasis without architectural changes, but we only identified this need through user check-ins; the pathway from ``the summary makes my day look inactive'' to ``identify a sensor imbalance'' to ``craft a targeted prompt'' requires domain experience and user-derived insight. 

A second, unavoidable cause is limited sensing coverage itself, raising a design question interviews alone could not surface: given that sensing data will always be incomplete, how can a system still convey an accurate feeling of someone's day? 
One solution is to let the system learn a person's baseline sensor activity before making a judgment, as we did by adding longitudinal context (e.g., average step counts over the past 30 days) to the LLM's input, similar to prior ML models that learn longitudinal trends to output anomaly labels~\cite{10510320,jakkula2008anomaly,cejudo2025smart}.
This approach is effective for simple labels such as `active/inactive', but once the content becomes richer, as in narrative summaries, user perception becomes complex and harder to ``fine-tune.'' 
For example, O1 reacted to the cumulative effect of the summary containing so many sentences describing in-home activities, which together created an impression of inactivity even without ever using the label ``inactive.''
We therefore also gave the LLM a detailed account of O1's typical routine (e.g., that she walked her dog three times daily), instructing it to reference this expectation and note whether sensing data could confirm or disconfirm each instance.

A final solution was letting older adults comment on the summary directly.
On the day O1 was sick, sensing data alone could not have captured this, but a brief self-reported message resolved it entirely, pointing to the broader potential of self-report when passive data is shared.
Beyond direct system messaging, conversational agents may offer a more user-friendly check-in channel for populations such as OAs \cite{mathur2022collaborative,mahmood2025voice,zubatiy2021empowering,yang2024talk2care}.
Future work should examine how OAs, aware that summaries may be shared, choose what to report via such channels.
Self-reporting, however, introduces its own limitation: when an OA's account conflicts with the sensors, whose version should reach RFMs? 
We discuss this power dynamic in the next section.

\subsection{An Open Book on Whose Terms?: Power Dynamics, Privacy, Autonomy, and Obligation in AI-Mediated Family Sharing}

The summaries served not only a practical purpose, but also constituted a form of AI-mediated communication (AI-MC): interpersonal communication in which an AI system modifies, augments, or generates content to achieve communicative and relational goals \cite{hancock2020ai}. Despite high compliance and engagement, this mediation introduced new obligations; in our case, a new expectation from O1 that F1 check the summary regularly, showing how a new communication ``actor'' can introduce implicit obligations into an established relationship.

AI-MC can also make existing communication gaps more visible. O1 and F1's low-key style, shaped partly by F1's demanding schedule, meant O1 was not always aware of what he cared about or worried over. She repeatedly expressed concern about sharing certain heart-rate patterns with him, not from discomfort disclosing it, but from uncertainty about how he would interpret it. This concern aligns with Communication Privacy Management theory~\cite{petronio2002boundaries,petronio2010communication}, in which conflict, or ``boundary turbulence,'' arises not from a violation of secrecy but from a mismatch between co-owners' expectations about how shared information will be received, explaining why O1's concern centered less on whether to share, and more on how it would be interpreted.
Back in 2010, Kirk et al. examined home video communication between generations, framing closeness-seeking as governed by a ``moral order of the home'': who has the right to look at whom, and what obligations that creates~\cite{kirk2010home}. 
AI-mediated sharing raises this question anew: our older adults described themselves as ``an open book,'' but on whose terms? 

A new AI-MC channel also raises a persistent question: does it replace more ``meaningful'' communication? 
Concretely, would our dyad stop calling or emailing once the summary began? 
Neither party did; the summary did not alter their underlying low-key communication style, consistent with social theories suggesting that adult child-parent relationships are structurally durable and persist across the lifespan~\cite{silverstein1997intergenerational,maine1861ancient,hogerbrugge2015transitions,suitor2011intergenerational}, with more continuity than change particularly in later life, where they are characterized more by continuity than change~\cite{hogerbrugge2015transitions,suitor2011intergenerational}.
As a result, new technologies tend to be absorbed into existing household routines rather than transforming them outright according to domestication theory~\cite{Silverstone1996DesignAT,soraa2021social}.

This raises a further question: what actually counts as ``meaningful'' communication?
Many might define it as visits or phone calls, not simply glancing at a summary~\cite{yuan2016they}. 
But should we be concerned that such tools are not driving more contact, when that contact might simply reflect worry, or should we instead think creatively about the new connections these tools can spark? 
F1 offered evidence for the latter: details he would otherwise never have known, such as events from the community calendar, gave him something new to discuss with O1. 
As summaries move beyond simply confirming a loved one is safe, we must consider how they will reshape communication through this new medium. 

\subsection{Design Implications}

Our findings point to several concrete implications for future AI-mediated sharing in complex socio-technical contexts.

\subsubsection{Aligning Users' Perceptions of Summaries with LLM Interpretation}
Beyond factual evaluation, future summaries should incorporate additional forms of techniques.
First, systems should proactively audit for \textbf{modality imbalance}, including \textbf{sensor coverage} (e.g., the number of sensors involved across different activities and locations) and \textbf{data volume} (e.g., the volume of sensor data generated during a typical event or within a fixed time frame). 
Second, systems should incorporate a user's \textbf{baseline activity} when summarizing subjective labels such as ``active/inactive''. 
This baseline should include both a \textbf{data baseline} (e.g., average steps over the past 30 days) and a \textbf{descriptive baseline} (e.g., what a typical day's routine looks like). Third, systems should involve \textbf{users' subjective feedback} in the evaluation process, particularly when content is being shared. 
For example, letting users flag important activities (e.g., leaving the apartment) they want tracked and shared with others, and surface the aspects users care most about when it comes to how they are portrayed, such as active/inactive or indoor/outdoor framing for O1. 
To achieve this, rather than weighting all modalities equally, systems should use multi-agent architectures like GLOSS, which support code generation and function calling to customize modality weighting. 
Finally, for real-world deployment with imperfect sensing data, systems should always give users a way to \textbf{comment on the content} themselves.

\subsubsection{Respecting Existing Power Dynamics: Enrichment over Replacement} 
Sharing AI-generated summaries should respect a dyad's existing social relationship. For intergenerational relationships, for example, Silverstein et al. identify five categories: tight-knit, sociable, intimate-but-distant, obligatory, and detached \cite{silverstein1997intergenerational}. 
Researchers should \textbf{respect these existing relationships}, particularly given their durability and stability, rather than attempting to reshape them. 
Overall, such shared media should be designed not as a replacement for existing communication, but as a means of \textbf{enrichment}, and thus contact frequency alone should not be treated as the success metric.

After introducing a new medium, however, designers should \textbf{proactively track resulting changes in communication}, focusing on \textbf{new obligations and expectations}, as well as \textbf{mismatches} in those obligations and expectations between different parties. More concretely, understanding how different stakeholders perceive and react to the same content differently and why to help guide more cautious design choices that avoid unexpected conflict.

In this study, we examined these tensions and implications within a relatively low-stakes OA-RFM relationship, but we hope our findings also surface insights relevant to higher-stakes dyads, such as an OA with dementia, where power dynamics are even more sensitive. 
Designing future AI-mediated data sharing for existing social relationships requires looking beyond accuracy or effectiveness alone, toward the underlying power dynamics, communication shifts, and new obligations such sharing introduces.

\subsection{Limitations}
Our work has several limitations. 
First, our relatively small sample size limits generalizability, and results should be interpreted with appropriate caution: although we deliberately selected interview days representing both full and partial data, all example summaries were drawn from a single older adult's data. 
Many participants immediately related the summaries to their own family members, but we still need to weight findings from these interviews with the caveat that reactions were to someone else's data, giving greater weight to the case study.
Second, our deployment involved only a single OA-RFM dyad; we therefore frame this component as a case study rather than a full-scale deployment. Although we described this dyad's communication style and living situation in detail, future work should examine a broader range of contexts.
Finally, we aimed to use summaries as an evolving artifact rather than evaluate their accuracy, thus our accuracy assessment relied only on researcher annotation and O1's own reflections, which may not capture every inaccuracy. 
O1, however, reviewed her summary almost daily and consistently reported inaccuracies as they occurred, whether through messages and emails or directly during interviews.
This consistency suggests her self-reports offer a reasonably reliable account of the summaries' accuracy.

\section{Conclusion} 
\seclabel{conclusion}
In this work, we investigated how AI-generated summaries of multi-modal sensing data can support sharing daily life between older adults (OAs) and remote family members (RFMs). Following a research-through-design approach, we began with formative interviews with 12 RFMs and 2 older adults, using early LLM-generated summaries to identify key needs, then redesigned the summary and deployed it over three months with a dyad and conducted regular check-ins and system iteration. The summaries became part of the dyad's morning routine during the three-month deployment, with both participants accessing more than 90\% of delivered summaries.
Yet our longitudinal, in-the-wild study surfaced tensions interviews alone could not reveal, tensions rooted not in factual errors, but in how technically accurate summaries could still misrepresent lived experience, and in how a new shared channel could surface pre-existing misalignments in a family's communication style. Together, these findings suggest that designing AI-mediated life-log sharing requires navigating the ongoing negotiation between what a system can accurately report and how that report is felt, interpreted, and acted upon by the people who receive it.

\ifanonymized
 \relax
\else
\fi 


\bibliographystyle{plain}	

\bibliography{bibs/local}

\end{document}